\documentclass[twocolumn,groupedaddress]{revtex4-1}
\usepackage{amsmath, bm, physics,amssymb}
\usepackage{color}
\usepackage{graphicx}
\usepackage{siunitx}
\graphicspath{{picFolder/}}

\usepackage[whole]{bxcjkjatype} 
\usepackage[colorlinks=true,urlcolor=blue,citecolor=blue,linkcolor=blue,breaklinks=true]{hyperref}
\input glyphtounicode
\newcommand{\smrm}[1]{_{\mathrm{#1}}}
\newcommand{\uprm}[1]{^{\mathrm{#1}}}

\begin{document}
\title{Emergent inductors in non-helical magnets}
\author{Takahiro Anan}
\affiliation{Department of Applied Physics, The University of Tokyo, Hongo, Tokyo,
  113-8656, Japan}
\author{Takahiro Morimoto}
\affiliation{Department of Applied Physics, The University of Tokyo, Hongo, Tokyo,
  113-8656, Japan}
\date{\today}
\begin{abstract}
The emergent inductor, which is a concept of an inductor employing quantum mechanics on helical magnets, has been studied actively from both theoretical and experimental aspects. Interestingly, finite inductance has been observed not only in spiral magnetic phases but also across various other magnetic phases although the underlying mechanism behind emergent inductance in non-helical magnets remains unresolved. In this study, we broaden the concept of emergent inductance to encompass non-helical magnets and establish a comprehensive formalism of the emergent inductance generated by magnetization dynamics. Using a diagrammatic approach, we derive the linear response of current density mediated by magnetization dynamics and formulate impedance and inductance for general magnetic materials. We reveal that the emergent inductance is composed of two distinct components, each contributing positively or negatively. Notably, the positive inductance arises from Ohmic dissipation including the effect of the emergent electric field (Berry phase in real space) while the negative inductance arises from the polarization of itinerant electrons (Berry curvature in $k$-$t$ space). We also apply the present method to quasi 1D models and show the numerical results of the inductance, the spin dynamics, the impedance, and the Q-value.
\end{abstract}
\maketitle

\section{Introduction}

Inductors are fundamental electronic components that exhibit frequency-dependent impedance and are used in electrical circuits as transformers or noise filters and so on.
However, the working principle of the prevailing inductor, which is constructed using a coil, follows classical electromagnetism, and thus the inductance of the coil is proportional to the cross section of the coil. 
This relationship poses an obstacle to miniaturization in electrical circuits.

\begin{figure}
  \includegraphics[width=\linewidth]{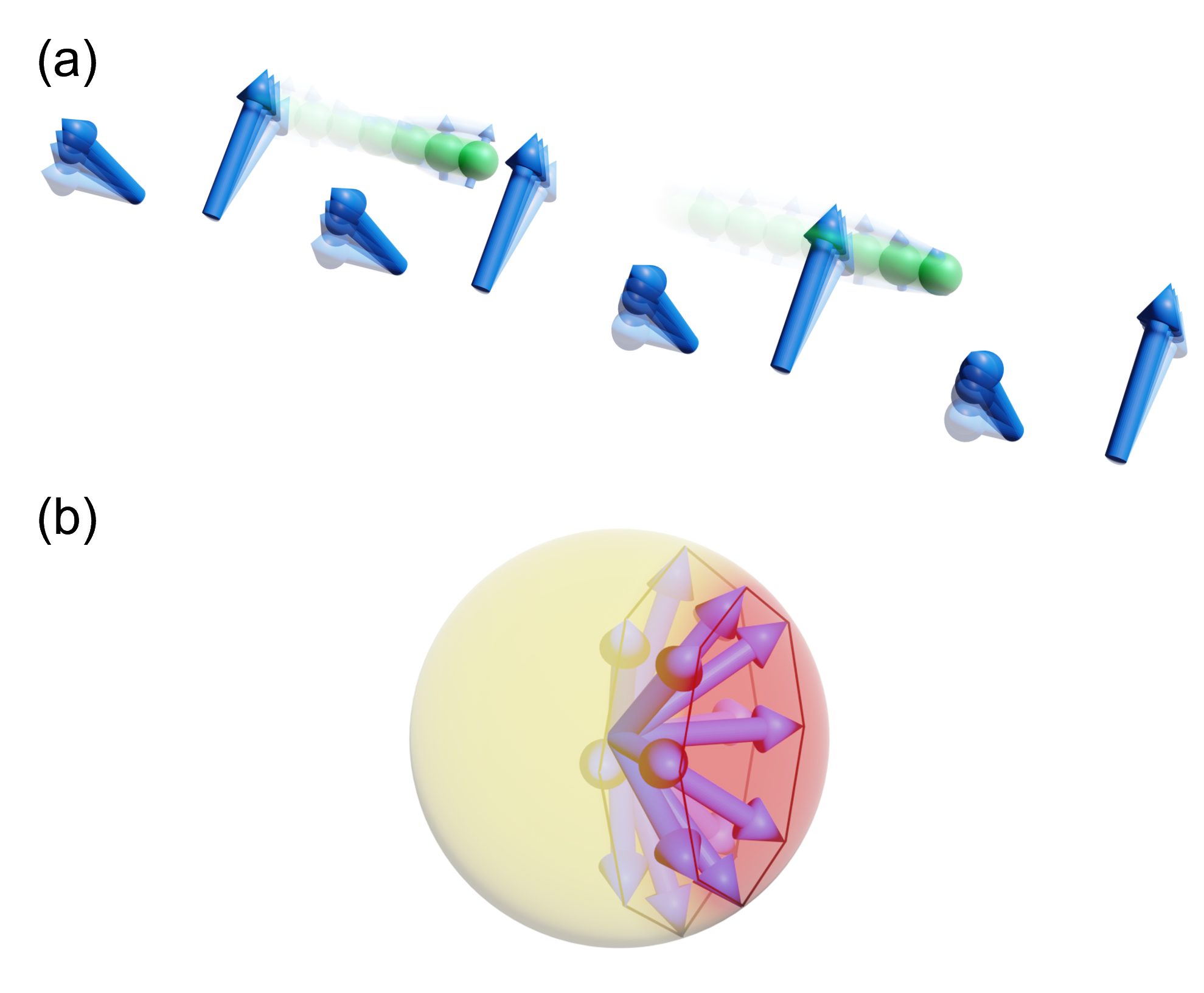}
  \caption{(a) Schematic picture of the emergent inductance in non-helical magnets.  (b) Schematic picture of the emergent electric field which is described as the time variation of the Berry phase induced by the spin configuration. The Berry phase is given as a solid angle surrounded by the local spins. In helical magnets, the emergent electric field is associated with the dynamics of the magnetization perpendicular to the co-plane of the helical spins.}
  \label{fig:schematic}
\end{figure}

In recent years, a novel concept of an inductor employing quantum mechanics on helical magnets has been proposed~\cite{Nagaosa2019}. 
This concept, referred to as the emergent inductor, has been studied actively from both theoretical and experimental aspects~\cite{Yokouchi2020,Kurebayashi2021,Kitaori2021,Kitaori2023,Kitaori2024}. 
When a current flows in helical magnets, spin transfer torque~\cite{Berger1996,Slonczewski1996} induces magnetization perpendicular to the co-plane of the helical spins (Fig.~\ref{fig:schematic}(b)), and the dynamics of this magnetization in turn drives the current by the spin motive force~\cite{Stern1992,Sakai2014}. 
The spin-motive force in a helical magnet is interpreted as the emergent electric field originating from the dynamical Berry phase in real space, which is caused by the helical spin texture. 
This inductance, unlike traditional inductors, has been shown to be inversely proportional to the sample's cross section, thereby overcoming the obstacle to miniaturization mentioned earlier.

The emergent inductance in helical magnets is also derived by a more quantum mechanical approach using the Green's function methods in Ref.~\cite{Kurebayashi2021}.
They introduce the phenomenological Lagrangian in terms of the magnetization mode and the phason mode that are the collective coordinates of the helical spin texture~\cite{Tatara2004} and evaluate the linear response mediated by these modes.
They rederive the positive inductance originating from the magnetization modes as shown in Ref.~\cite{Nagaosa2019}, and also find that the phason mode of helical magnets gives rise to the negative inductance in the presence of finite pinning potential.

The theories of the emergent inductance in helical magnets have spurred researchers  to search for materials that exhibit this property,
and a few measurements of the emergent inductance have already been reported~\cite{Yokouchi2020,Kitaori2021,Kitaori2023,Kitaori2024}.
The first reported material was $\mathrm{Gd}_3\mathrm{Ru}_4\mathrm{Al}_{12}$~\cite{Yokouchi2020}, followed by $\mathrm{YMn}_6\mathrm{Sn}_6$, which exhibits emergent inductance at room temperature~\cite{Kitaori2021}. 
Furthermore, the inductances of the terbium-doped system $\mathrm{Y}_{1-x}\mathrm{Tb}_x\mathrm{Mn}_6\mathrm{Sn}_6$ were measured for the purpose of investigating the doping dependence of the inductance~\cite{Kitaori2023},
and the frequency dependence of the emergent inductance was investigated in $\mathrm{Tb}_5\mathrm{Sb}_3$~\cite{Kitaori2024}.

It is worth noting that the finite inductance has been observed not only in spiral phases but also in other phases~\cite{Yokouchi2020,Kitaori2021,Kitaori2023,Kitaori2024}.
Nevertheless, the previous theoretical research on the emergent inductance originating from the non-colinear magnetization~\cite{Nagaosa2019,Kurebayashi2021} was specialized for helical magnets, and as a result, experimental studies have been conducted only on materials having helical phases. 
Thus, theoretical elucidation of the emergent inductance in non-helical magnets will expand the scope of material search.
Furthermore, it has been demonstrated that the frequency range of emergent inductors is limited by the pinning potential of the magnets~\cite{Nagaosa2019, Kurebayashi2021}.
Therefore, emergent inductors that work at high frequencies are likely to be discovered in non-helical magnets whose pinning potential is larger than that of magnetic helices.
In this context, there is a demand for the theory for emergent inductance in general magnetic structures which clarifies the origin of emergent inductance in non-helical magnets and provides a guiding principle for material search of more efficient inductors.

In the present paper, we study the general formalism of the emergent induction arising from the dynamics of magnetization and analyze its physical meaning (Fig.~\ref{fig:schematic}(a)).
We derive the linear response of the current density mediated by the dynamics of magnetization using a diagrammatic approach and formulate the impedance and inductance for general magnetic materials.
We show that the inductance is composed of two components that contribute positively and negatively to the inductance, like a coil and a capacitor, respectively.
We also show that the positive inductance arises from Ohmic dissipation including the effect of the emergent electric field (Berry phase in real space), while the negative inductance arises from the polarization of the itinerant electrons (Berry curvature in $k$-$t$ space).
Furthermore, we apply the present method to several quasi 1D models of non-helical magnets and numerically study behaviors of the inductance, the spin dynamics producing finite inductance, the impedance, and the Q-value.
Finally, we discuss the candidate materials for emergent inductors.

The rest of the paper is organized as follows.
In Sec.~\ref{sec:formalism}, we present the formalism of the emergent inductance and the impedance arising from the dynamics of magnetization.
In Sec.~\ref{subsec:model}, we introduce the model Hamiltonian of the itinerant electron coupled with localized spins and transform them to the rotated frame which allows us to clearly understand the physical picture of how the magnetization fluctuation couple with the spins of itinerant electrons.
In Sec.~\ref{subsec:linearResponse}, using a diagrammatic technique, we derive the general expression of the linear response of the current density induced by the monochromatic AC electric field.
In Sec.~\ref{subsec:sigmaChi}, we derive the response functions of the itinerant electron system up to the first order in frequency $\omega$.
In Sec.~\ref{subsec:physicalMeaning}, we discuss the physical meaning of these response functions.
In Sec.~\ref{subsec:GreenFunction}, we derive the Green's function of the spin Hamiltonian and obtain the conductivity mediated by the dynamics of magnetization, 
with which we obtain the impedance and the inductance.
In Sec.~\ref{sec:InductanceAndImpedance}, we present the application of the present method for quasi 1D models.
We introduce the 1D models in Sec.~\ref{subsec:1Dmodels} and present the numerical results of the inductance (Sec.~\ref{subsec:Inductance}), the spin dynamics (Sec.~\ref{subsec:SpinDynamics}), the impedance (Sec.~\ref{subsec:Impedance}) and Q-value (Sec.~\ref{subsec:QValue}).
In Sec.~\ref{sec:Discussion}, we give a brief discussion of candidate materials and the comparison with previous studies.

\section{Formalism}\label{sec:formalism}
In this section, we present the formalism of the emergent inductance for general magnetic materials.
Toward this goal, we derive the current response mediated by the dynamics of magnetization.
We introduce the general tight-binding model and express it in the rotated frame in which all the localized spins are aligned along the $z$-axis (Sec.~\ref{subsec:model}).
In the rotated frame, the Hund coupling divides into the effective magnetic field and the coupling between the magnetization fluctuation and the itinerant electrons, which allows us to clearly understand the physical picture of how the magnetization fluctuation couples with the spins of itinerant electrons.
In this rotated frame, we derive the linear response of the spin dynamics-mediated current induced by the monochromatic AC electric field using a diagrammatic approach (Fig.~\ref{fig:FeynmanDiagram}).
In Sec.~\ref{subsec:linearResponse}, we derive the response functions of the itinerant electron system under the effective magnetic field from the localized spins up to the first order in frequency $\omega$.
Specifically, we derive the response functions of the current density induced by the dynamics of the magnetization $\sigma(\omega)$ and the response functions of the magnetic moment of itinerant electrons induced by the AC field $\chi(\omega)$.
In Sec.~\ref{subsec:GreenFunction}, we derive the impedance and inductance.
We derive the Green's function of the spin Hamiltonian $D(\omega)$ and obtain the conductivity mediated by the dynamics of magnetization $\Sigma(\omega)$ by using $\sigma(\omega)$, $\chi(\omega)$ and $D(\omega)$.
We derive the impedance and the inductance from the conductivity $\Sigma(\omega)$.
Furthermore, we discuss the physical meaning of the two components appearing in both $\sigma(\omega)$ and $\chi(\omega)$ in Sec.~\ref{subsec:physicalMeaning}, and show these two components contribute positively and negatively to the inductance, like a coil and a capacitor, respectively in Sec.~\ref{subsec:GreenFunction}.

\subsection{Model}\label{subsec:model}
We consider the general tight-binding model describing the itinerant electron coupled with localized spins without spin-orbit coupling (SOC) focusing on the effect of the dynamics of magnetization as following:
\begin{align}
  H & =H\smrm{itinerant}+H\smrm{Hund}+H\smrm{m}, \\
  H\smrm{itinerant} & =-\sum_{i,j,\tau} t_{j,\tau}c_{(i,j)}^\dagger \sigma_0 c_{(i,j)+\tau}+h.c.,\label{eq:HelLab} \\
  H\smrm{Hund} & =-\sum_{i,j} \frac{J_j\uprm{H}}{2}\bm{S}_{(i,j)} \cdot c_{(i,j)}^\dagger \bm{\sigma }c_{(i,j)},\label{eq:HcLab} \\
  H\smrm{m} & =-\sum_{i,j,\tau,\mu,\nu} J_{j,\tau}^{\mu\nu}S_{(i,j)}^\mu S_{(i,j)+\tau}^\nu-\sum_{i,j}\bm{B}_j\cdot \bm{S}_{(i,j)}.\label{eq:HmLab}
\end{align}
where $i$ is the index of the magnetic unit cell, $j$ is the index of the site in a unit cell, 
$c_{(i,j)}$ is the annihilation operator of the itinerant electron at the site $(i,j)$,
$\bm{ S}_{(i,j)}$ is the spin operator at the site $(i,j)$,
and $\sigma_{\mu=x,y,z}$ is the Pauli matrix.
$t_{(i,j),\tau}$ represents the hopping amplitude between the site $(i,j)$ and $(i,j)+\tau$ if $\tau\neq0$ with $\tau$ being a bond vector, and the on-site potential if $\tau=0$.
As for $H\smrm{m}$, $J_{j,\tau}^{\mu\nu}$ represents various coupling constants or potential depending on the value of $\tau$, $\mu$ and $\nu$.
For the case of $\tau\neq 0$, $J_{j,\tau}^{\mu\nu}$ represents the exchange coupling between the localized spins at site $(i,j)$ and $(i,j)+\tau$ if $\mu=\nu$, Dzyaloshinsky-Moriya interactions or anisotropic interactions if $\mu\neq \nu$,
and, for the case of $\tau=0$, $J_{j,\tau}^{\mu\nu}$ represents the pinning potential if $\mu=\nu$ and the anistropy if $\mu\neq \nu$.
$J_j\uprm{H}$ represents the exchange coupling between the itinerant electron and the localized spin at site $(i,j)$.
$\bm{B}_j$ represents the external magnetic field or effective magnetic field by the exchange coupling with other localized spins acting on the localized spin at site $(i,j)$.
We note that the effect of SOC on the emergent inductance has already been discussed in Refs.~\cite{Ieda2021,Yamane2022,Araki2023}.
In the present methods, we do not consider SOC in the itinerant electron system, focusing on the effect of the dynamics of nontrivial magnetic structures.
The treatment of the SOC in the present is discussed in Sec.~\ref{subsec:Comparison} and the numerical results are shown in Appendix~\ref{app:spinorbit}.

We assume that the spin system described by $H\smrm{m}$ realizes the classical spin configuration that is the ground state of $H\smrm{m}$
and discuss the dynamics where the spin configuration of the ground state is slightly fluctuated by an AC electric field.
Therefore, we move to the coordinate system that allows us to easily distinguish the fluctuation of the magnetization from the ground state.
Namely, we perform the coordinate transformation such that all the localized spins of the ground state are aligned along the $z$-axis.
Note that this procedure makes it clear the physical meaning of the subsequent calculation,  while one may also adopt a formulation without the coordinate transformation for deriving the emergent inductance, as detailed in Appendix~\ref{app:magnonFormalism}.
We denote the spin configuration of the ground state of the spin Hamiltonian $H\smrm{m}$ as $\{\bm{n}_j\},\bm{n}_j=(\sin\theta_j \cos\phi_j,\sin\theta_j \sin\phi_j,\cos \theta_j)\uprm{T}$ and assume that there is a finite gap to the first excited state.
We define a SO(3) rotation operator $T_j$ that rotates classical spin from up spin to $\bm{n_j}$ and the corresponding SU(2) operator $U_j$.
In other words, $T_j(0,0,1)\uprm{T}=\bm{n}_{j}$, and $U_j$ satisfies $(T_j\bm v )\cdot \bm{\sigma} =U_j (\bm v \cdot \bm{ \sigma}) U_j^\dagger$ for all $\bm{ v}\in \mathbb{R}^3$.
In the present paper, we adopt the following representation of $T_j$ and $U_j$:
\begin{align}
  T_j & = \left(
  \begin{array}{ccc}
      \cos \phi_j \cos \theta_j & -\sin \phi_j & \cos \phi_j \sin \theta_j \\
      \sin \phi_j \cos \theta_j & \cos \phi_j & \sin \phi_j \sin \theta_j \\
      -\sin \theta_j & 0 & \cos \theta_j \\
    \end{array}
  \right), \label{eq:T} \\
  U_j & =\begin{pmatrix}
           e^{-i\phi_j/2}\cos\frac{\theta_j}{2} & -e^{-i\phi_j/2}\sin\frac{\theta_j}{2} \\
           e^{i\phi_j/2}\sin\frac{\theta_j}{2} & e^{i\phi_j/2}\cos\frac{\theta_j}{2} \label{eq:U}
         \end{pmatrix}.
\end{align}
We define the spin operator and the electron operator in the rotated frame as $\bm{S}_{(i,j)}=T_j \tilde{\bm{S}}_{(i,j)}$ and $c_{(i,j)}=U_j \tilde{c}_{(i,j)}$.
Using Eqs.~\eqref{eq:T} and \eqref{eq:U}, we obtain the Hamiltonian in the rotated frame,
\begin{align}
  H\smrm{itinerant} & =-\sum_{i,j,\tau} t_{j,\tau}\tilde{c}_{(i,j)}^\dagger U_j^\dagger U_{j+\tau} \tilde{c}_{(i,j)+\tau}+h.c.,\label{eq:HelFerro} \\
  H\smrm{Hund} & =-\sum_{i,j} \frac{J_j\uprm{H}}{2}\tilde{\bm{S}}_{(i,j)} \cdot \tilde{c}_{(i,j)}^\dagger \bm{\sigma }\tilde{c}_{(i,j)},\label{eq:HcFerro} \\
  H\smrm{m} & =-\sum_{i,j,\tau,\mu,\nu} \tilde{J}_{j,\tau}^{\mu\nu}\tilde{S}_{(i,j)}^\mu \tilde{S}_{(i,j)+\tau}^\nu-\sum_{i,j}\tilde{\bm{B}}_j \cdot \tilde{\bm{S}}_{(i,j)},\label{eq:HmFerro}
\end{align}
where $\tilde{J}_{j,\tau}=T_j J_{j,\tau}T_{j+\tau}\uprm{T}$ and $\tilde{\bm{B}}_j=T_j^\mathrm{T} \bm{B}_j$.
Since the ground state of the spin Hamiltonian $H\smrm{m}$ is considered the ferromagnetic state in the rotated frame,
$H\smrm{m}$ in the magnon representation does not contain any single magnon term.
Additionally, we neglect the higher order terms of the magnon operator.
We approximate $H\smrm{Hund}$ using $\tilde{S}_{(i,j)}^z \tilde{c}_{(i,j)}^\dagger \sigma_z \tilde{c}_{(i,j)}\simeq \hbar S\tilde{c}_{(i,j)}^\dagger \sigma_z \tilde{c}_{(i,j)}$ whose right hand side is no longer the coupling but serves as the effective magnetic field in $H\smrm{itinerant}$.
Finally, performing the Fourier transformation
$\tilde{c}_{(i,j)}=\frac{1}{\sqrt{N}}\sum_{\bm{k}}e^{i\bm{k} \cdot\bm{r}_{(i,j)}}\tilde{c}_{\bm{k},j},\tilde{S}_{(i,j)}=\frac{1}{\sqrt{N}}\sum_{\bm{k}}e^{i\bm{k} \cdot\bm{r}_{(i,j)}}\tilde{S}_{\bm{k},j}$ with $\bm{r}_{(i,j)}$ being a position vector, we obtain the Hamiltonian in the momentum space,
\begin{align}
  H\smrm{el}= & -\sum_{\bm{k},j,\tau} [t_{j,\tau}e^{i\bm{k}\cdot \bm{r}_\tau}\tilde{c}_{\bm{k},j}^\dagger U_j^\dagger U_{j+\tau} \tilde{c}_{\bm{k},j+\tau}+h.c.]\nonumber \\
 & \qquad -\sum_{\bm{k},j}\frac{J_j\uprm{H}\hbar S}{2}\tilde{c}_{\bm{k},j}^\dagger \sigma_z \tilde{c}_{\bm{k},j}, \label{eq:HelandMagF} \\
  H\smrm{c}= & \frac{1}{\sqrt{N}}\sum_{\bm{k},\bm{q},j} \frac{-J_j\uprm{H}}{2}[\tilde{S}_{\bm{q},j}^x \cdot \tilde{c}_{\bm{k}+\bm{q},j}^\dagger \sigma_x \tilde{c}_{\bm{k},j}\nonumber \\
  &\qquad\qquad\qquad+\tilde{S}_{\bm{q},j}^y \cdot \tilde{c}_{\bm{k}+\bm{q},j}^\dagger \sigma_y \tilde{c}_{\bm{k},j}], \\
  H\smrm{m}= & -\sum_{\bm{k},j,\tau,\mu,\nu} \tilde{J}_{j,\tau}^{\mu\nu}\tilde{S}_{-\bm{k},j}^\mu \tilde{S}_{\bm{k},j+\tau}^\nu e^{i\bm{k} \cdot \bm{r}_\tau} -\sum_{j}\tilde{\bm{B}}_j\cdot \tilde{\bm{S}}_{0,j}\label{eq:Hmkspace}.
\end{align}
In this representation, it is clear that $H\smrm{c}$ is the coupling between the magnetization fluctuation $\tilde{S}_{\bm{q},j}^{\mu=x,y}$ and the itinerant electrons.

\subsection{Linear response}\label{subsec:linearResponse}
\begin{figure}
  \includegraphics[width=\linewidth]{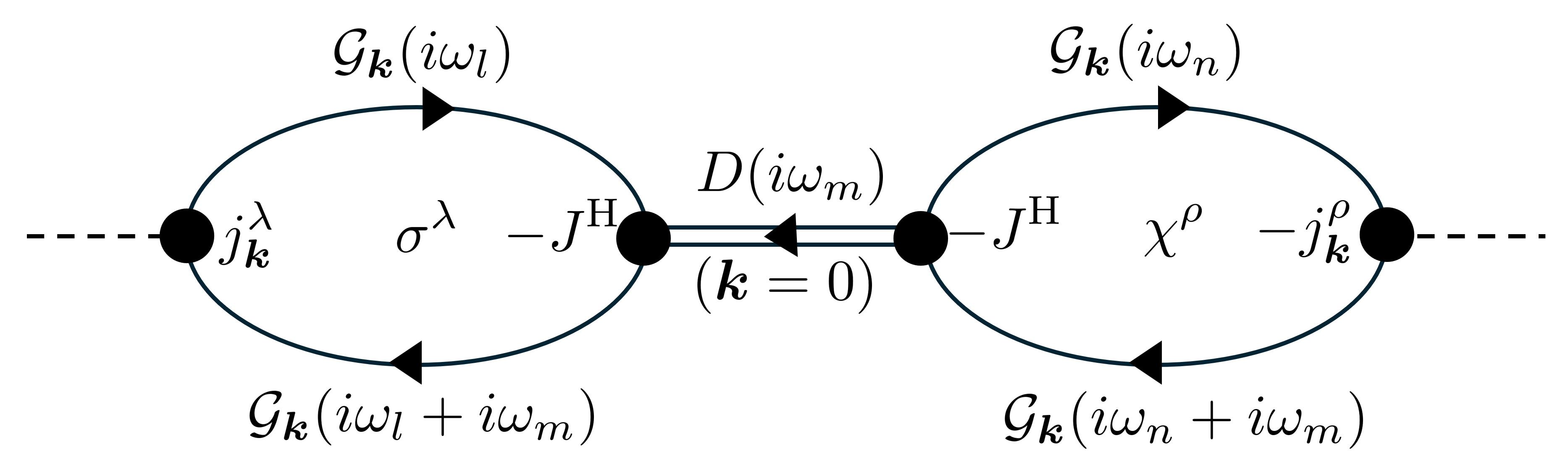}
  \caption{Feynman diagram for the current response responsible for the emergent inductance. The solid line represents the Green's function of the itinerant electron, and the double solid line represents the Green's function of the localized spin. The dashed line represents the AC electric field. The left bubble diagram represents the response of current density induced by the dynamics of magnetization, while the right bubble diagram represents the response of the expectation value of the magnetic moment of itinerant electrons induced by the AC field.}
  \label{fig:FeynmanDiagram}
\end{figure}
We consider the linear response of the current density induced by the monochromatic AC electric field, $j^\mu (\omega)=j^\mu \smrm{dc}+j^\mu \smrm{m}(\omega)=\sum_\nu(\sigma \smrm{dc}^{\mu\nu}+\Sigma^{\mu\nu} (\omega))E^\nu(\omega)$, by extending the method in Ref.~\cite{Kurebayashi2021} to address magnetic materials with general spin structures.
Here, $j\smrm{dc}$ represents the contribution solely from the electron system, while $j\smrm{m}$ represents the contribution of currents mediated by the excitation of magnetization.
Considering $H\smrm{c}$ and $-jA$ as perturbations, we describe $j\smrm{m}$ in imaginary time formalism as follows:
\begin{align}
  j\smrm{m}(\tau) & =\frac{1}{V} \int d\tau_1 d\tau_2 d\tau_3 \Bigl \langle\mathcal{T} \sum_{\bm{k},j,j'}j_{\bm{k},j,j'} \tilde{c}_{\bm{k},j}^\dagger(\tau) \tilde{c}_{\bm{k},j'}(\tau)\nonumber \\
 & \times[ \frac{1}{\sqrt{N}}\sum_{\bm{k},\bm{q},j,\mu} \frac{-J_j\uprm{H}}{2}\tilde{S}_{\bm{q},j}^\mu (\tau_1) \cdot \tilde{c}_{\bm{k}+\bm{q},j}^\dagger (\tau_1)  \sigma_\mu \tilde{c}_{\bm{k},j} (\tau_1)  ]\nonumber \\
 & \times [\frac{1}{\sqrt{N}}\sum_{\bm{k},\bm{q},j,\mu} \frac{-J_j\uprm{H}}{2}\tilde{S}_{\bm{q},j}^\mu (\tau_2)  \cdot \tilde{c}_{\bm{k}+\bm{q},j}^\dagger (\tau_2)  \sigma_\mu \tilde{c}_{\bm{k},j} (\tau_2) ]\nonumber \\
 & \times \sum_{\bm{k},j,j'}(-j_{\bm{k},j,j'} )\tilde{c}_{\bm{k},j}^\dagger (\tau_3)  \tilde{c}_{\bm{k},j'} (\tau_3)A(\tau_3)\Bigr \rangle\label{eq:pathInt}
\end{align}
where $\tilde{c}_{\bm{k},j'}(\tau)=e^{(H\smrm{el}+H\smrm{m})\tau}\tilde{c}_{\bm{k},j'}e^{-(H\smrm{el}+H\smrm{m})\tau}$ and $\tilde{S}_{\bm{q},j}^\mu (\tau)=e^{(H\smrm{el}+H\smrm{m})\tau}\tilde{S}_{\bm{q},j}^\mu e^{-(H\smrm{el}+H\smrm{m})\tau}$ are the interaction pictures of the electron operator and the spin operator, respectively, $V$ is the volume (length for 1D case and area for 2D case) of the sample, and the current velocity is given as $j_{\bm{k}}^\mu=-\frac{e}{\hbar}\partial_{\bm{k}_\mu}\mathcal{H}\smrm{el}(\bm{k})$ with $\mathcal{H}\smrm{el}(\bm{k})$ being the coefficient matrix of $H\smrm{el}$ in $c_{\bm{k},j}$ representation.
Finally, we obtain the linear response kernel in frequency space by performing the Fourier transformation,
\begin{align}
  \Sigma^{\lambda\rho}(i\omega_m) & =\sum_{i,j,\mu,\nu}\sigma_{(i,\mu)}^\lambda(i\omega_m) D_{(i,\mu),(j,\nu)}(i\omega_m)\nonumber \\
 & \qquad \times \frac{-J_j\uprm{H}}{2} \chi_{(j,\nu)}^\rho (i\omega_m) \frac{1}{-\omega_m}\label{eq:Sigma} \\
  \sigma_{(i,\mu)}^\lambda(i\omega_m) & \equiv\frac{1}{\beta V}\sum_{\bm{k},\omega_l}\mathrm{Tr}[j_{\bm{k}}^\lambda \mathcal{G}_{\bm{k}}(i\omega_l+i\omega_m)\nonumber \\
 & \qquad \qquad\qquad \times (\Pi_i\sigma_\mu)\mathcal{G}_{\bm{k}}(i\omega_l)]\frac{-J_i\uprm{H}}{2} \label{eq:sigmaIMu} \\
  \chi_{(j,\nu)}^\rho (i\omega_m) & \equiv \frac{1}{\beta N}\sum_{\bm{k},\omega_n}\mathrm{Tr}[ (\Pi_j\sigma_\nu)\mathcal{G}_{\bm{k}}(i\omega_n +i\omega_m)\nonumber \\
 & \qquad \qquad\qquad \times (-j_{\bm{k}}^\rho)\mathcal{G}_{\bm{k}}(i\omega_n)]. \label{eq:chiJNu}
\end{align}
Here, $[\Pi_{j}]_{a,b}=\delta_{a,b}\delta_{j,a}$ is the matrix of the projection operator to the site $j$ in a magnetic unit cell, where $a,b$ are labels for the sublattice degrees of freedom.
$\mathcal{G}_{\bm{k}}(i\omega)$ is the Green's function of the itinerant electron with the indices of sites in a magnetic unit cell and spin,
and $D(i\omega)$ is the Green's function of the magnon at $\bm{q}=0$ with the indices of the sites and the direction of perturbation $x,y$.
Figure~\ref{fig:FeynmanDiagram} is the Feynman diagram of the response of the $j\smrm{m}$ induced by the AC electric field.
Note that the non-zero Green's functions are composed of $\bm{q}=0$ operator in Eq.~\eqref{eq:pathInt}.
Also, while there can appear diamagnetic contributions (vertices of the order $A$ in the diagram) in principle, those contributions vanish as shown in Eq.~\eqref{eqApp:diamag} in Appendix \ref{app:Cderivation}. 
We will evaluate $\sigma_{(i,\mu)}(\omega), \chi_{(j,\nu)} (\omega), D_{(i,\mu),(j,\nu)}(\omega) $ below.

\subsection{$\sigma,\chi$, Fermi surface terms and interband effect terms}\label{subsec:sigmaChi}
In this section, we derive the response functions of the current density $j\smrm{m}^\lambda$ induced by the fluctuation of magnetization (Eq.~\eqref{eq:sigmaIMu}) and the response functions of the expectation values of the Pauli matrix $\sigma_{\nu=x,y}$ of itinerant electrons induced by the AC field (Eq.~\eqref{eq:chiJNu}).

First, we derive $\sigma_{(i,\mu)}^\lambda(\omega)$ that is the response function of the current density $j\smrm{m}^\lambda(\omega)$ from the fluctuation of the magnetization $\tilde{S}_i^\mu$ at site $i$.
By performing analytic continuation of $i\omega_m$ and $i\omega_l$ in Eq.~\eqref{eq:sigmaIMu}, we obtain the response function of the current density $j\smrm{m}^\lambda(\omega)$.
Assuming that the frequency $\omega$ is sufficiently smaller than the energy scale of $H\smrm{el}$, we extract the $\omega$ linear term and obtain
\begin{align}
  \sigma_{(i,\mu)}^\lambda & (\omega)    =i\hbar \omega C_{\mathrm{O}(i,\mu)}^\lambda\frac{-J_i\uprm{H}}{2}+i\hbar \omega C_{\mathrm{I}(i,\mu)}^\lambda\frac{-J_i\uprm{H}}{2}\label{eq:sigmaIMuWithC} \\
  C_{\mathrm{O}(i,\mu)}^\lambda & =\frac{\tau}{\hbar }\sum_{u}\int_{\mathrm{BZ}} \frac{d\bm{k}}{(2\pi)^d} \mel{u}{{j}_{\bm{k}}^\lambda}{u}\mel{u}{\Pi_i\sigma_\mu}{u}f_0'(\xi_{\bm{k},u}) \label{eq:CO} \\
  C_{\mathrm{I}(i,\mu)}^\lambda & =2\mathrm{Im}\sum_{u\neq v}\int_{\mathrm{BZ}} \frac{d\bm{k}}{(2\pi)^d}\frac{\mel{u}{{j}_{\bm{k}}^\lambda}{v}\mel{v}{\Pi_i\sigma_\mu}{u}}{(\xi_{\bm{k},u}-\xi_{\bm{k},v})^2}f_0(\xi_{\bm{k},u})  \label{eq:CI}
\end{align}
where $f_0(\xi)\equiv(e^{\beta \xi}+1)^{-1}$ is the Fermi distribution function, $\xi_{\bm{k},u}=\varepsilon_{\bm{k},u}-\mu$, $\varepsilon_{\bm{k},u}$ is the eigenenergy of the band $u$ at $\bm{k}$, $\mu $ is the chemical potential, and $\tau$ is the relaxation time. 
For details of derivation, see Appendix \ref{app:Cderivation}.
As can be seen from Eq.~\eqref{eq:sigmaIMuWithC}, $\sigma_{(i,\mu)}^\lambda$ is written with two real numbers $C_{\mathrm{O}(i,\mu)}^\lambda$ and $C_{\mathrm{I}(i,\mu)}^\lambda$.
$C_{\mathrm{O}(i,\mu)}^\lambda$ is the Fermi surface term that only has contributions from the Fermi surface because of $f_0'(\xi_{\bm{k},u})$.
This term is proportional to the relaxation time $\tau$ and a dissipative term.
In contrast, $C_{\mathrm{I}(i,\mu)}^\lambda$ is the interband effect terms that have contributions from the matrix element between the bands $u$ and $v$.
This term has a finite value even in insulating systems.
The Fermi surface terms are dominant in clean metals $\tau \gg \hbar/|\xi_u-\xi_v|$, while the interband effect terms are dominant in bad metals $\tau \ll \hbar/|\xi_u-\xi_v|$ or insulators.

Next, we evaluate $\chi_{(j,\nu)}^\rho(\omega)$ that is the response function of the expectation value of the Pauli matrix of itinerant electrons at site $j$, ($\ev{\Pi_j \sigma_{\nu=x,y}}$), induced by an AC electric field.
By comparing the Eqs.~\eqref{eq:sigmaIMu} and~\eqref{eq:chiJNu}, we obtain
\begin{align}
  \chi_{(j,\nu)}^\rho(\omega)=-i\hbar\omega v C_{\mathrm{O}(j,\nu)}^\rho+i\hbar \omega v C_{\mathrm{I}(j,\nu)}^\rho, \label{eq:chiJNuWithC}
\end{align}
where $v\equiv V/N$ is the volume of a unit cell.
Here, we use $C_{\mathrm{O}(j,\nu)}^\rho$ and $C_{\mathrm{I}(j,\nu)}^\rho$ in Eqs.~\eqref{eq:CO} and~\eqref{eq:CI}.

\subsection{Physical meaning of Fermi surface terms $C_{\mathrm{O}(i,\mu)}^\lambda$ and interband effect terms $C_{\mathrm{I}(i,\mu)}^\lambda$}\label{subsec:physicalMeaning}
In this section, we discuss the physical meaning of the Fermi surface terms $C_{\mathrm{O}(i,\mu)}^\lambda$ and the interband effect terms $C_{\mathrm{I}(i,\mu)}^\lambda$.
While $C_{\mathrm{O}(i,\mu)}^\lambda$ and $C_{\mathrm{I}(i,\mu)}^\lambda$ are derived from the diagrammatic approach, we can also derive $C_{\mathrm{O}(i,\mu)}^\lambda$ and $C_{\mathrm{I}(i,\mu)}^\lambda$ by using the Boltzmann transport theory and the Berry phase formula for the polarization, respectively.
In the following, using these approaches, we show that $C_{\mathrm{O}(i,\mu)}^\lambda$ represents the contribution from the emergent electric field in the limit of large Hund coupling $J\uprm{H}$ when no loop of bonds encloses emergent magnetic flux,
 and $C_{\mathrm{I}(i,\mu)}^\lambda$ represents the contribution of the time derivative of the polarization.

\subsubsection{Fermi surface terms $C_{\mathrm{O}(i,\mu)}^\lambda$}\label{subsubsec:CO}
We rederive $C_{\mathrm{O}(i,\mu)}^\lambda$ using the Boltzmann transport theory and discuss the physical meaning of $C_{\mathrm{O}(i,\mu)}^\lambda$.
The outline of the derivation and discussion is as follows.
First, we derive the $C_{\mathrm{O}(i,\mu)}^\lambda$ using the Boltzmann transport theory.
Second, we consider the limit of the large Hund coupling, and make the approximation that the response current $j_\mathrm{O}^\lambda$ 
is obtained from the spin-polarized Hamiltonian in the projected Hilbert space where the itinerant spins are aligned with the localized spins (denoted as $H_{\mathrm{el}\uparrow}$ in Eq.~\eqref{eq:HelUp}).
Finally, we show that the approximated $j_\mathrm{O}^\lambda$ is attributed to the emergent electric field.

We consider the Boltzmann equation of the distribution function $f(\xi_{\bm{k},u}(t))$,
\begin{align}
  -\frac{f-f_0}{\tau}=\frac{df}{dt}=\frac{df}{d \xi_{\bm{k},u}}\left(\frac{\partial \xi_{\bm{k},u}}{\partial \bm{k}}\cdot \dot{\bm{k}}+\frac{\partial \xi_{\bm{k},u}}{\partial t}\right). \label{eq:BoltzmannEq}
\end{align}
Using this Boltzmann equation, we can derive the current density $j_{\bm{k}}^\lambda$ as,
\begin{align}
  j_{\mathrm{O}}^\lambda & =\sum_u\int\smrm{BZ} \frac{d\bm{k}}{(2\pi)^d} \mel{u}{j_{\bm{k}}^\lambda}{u} f \nonumber \\
 & =-\tau \sum_u\int\smrm{BZ} \frac{d\bm{k}}{(2\pi)^d} \mel{u}{j_{\bm{k}}^\lambda}{u} \frac{\partial f}{\partial \xi_{\bm{k},u}}\left( \frac{\partial \xi_{\bm{k},u}}{\partial \bm{k}}\cdot \dot{\bm{k}} +\frac{\partial \xi_{\bm{k},u}}{\partial t}\right)\nonumber \\
 & \simeq -\tau \sum_u\int\smrm{BZ} \frac{d\bm{k}}{(2\pi)^d} \mel{u}{j_{\bm{k}}^\lambda}{u} \frac{\partial f_0}{\partial \xi_{\bm{k},u}}\left( \frac{\partial \xi_{\bm{k},u}}{\partial \bm{k}}\cdot \dot{\bm{k}} +\frac{\partial \xi_{\bm{k},u}}{\partial t}\right).
\end{align}
Here, we assume that the current density in equilibrium is zero in the second line and approximate the distribution function as $f\simeq f_0$ in the third line.
To derive the response current by the spin fluctuation, we consider the case that there are no external fields $\dot{\bm{k}}=0$ and obtain the current density as,
\begin{align}
  j_{\mathrm{O}}^\lambda =-\tau \sum_u\int\smrm{BZ} \frac{d\bm{k}}{(2\pi)^d} \mel{u}{j_{\bm{k}}^\lambda}{u} \frac{\partial f_0}{\partial \xi_{\bm{k},u}}\frac{\partial \xi_{\bm{k},u}}{\partial t}. \label{eq:COBoltzmann}
\end{align}

Now, we evaluate $\frac{\partial \xi_{\bm{k},u}}{\partial t}$ under the fluctuation of the localized spins.
The Hamiltonian of the itinerant electrons subjected to the effective magnetic field from the fluctuated localized spins is given as,
\begin{align}
  H\smrm{el}(t)= & H\smrm{el}+H\smrm{c}(t), \\
  H\smrm{c}(t)= & -\sum_{\substack{\bm{k},j, \\ \mu=x,y}}\frac{J_j\uprm{H}\hbar S}{2}\tilde{n}_{0,j}^\mu(t) \tilde{c}_{\bm{k},j}^\dagger \sigma_\mu \tilde{c}_{\bm{k},j}, \label{eq:HcTimeDependent}
\end{align}
where $\tilde{n}_{0,j}^\mu(t)$ is the $\bm{q}=0$ component of the fluctuation of the localized spin at site $j$ in rotated frame.
Therefore, the time derivative of the band energy of the itinerant electrons is given as,
\begin{align}
  \frac{\partial \xi_{\bm{k},u}}{\partial t} & =\mel{u}{\frac{\partial \mathcal{H}'\smrm{el}}{\partial t}}{u}\nonumber \\
 & =\sum_{j,\mu}\mel{u}{\Pi_j \sigma_\mu}{u}\frac{d \tilde{n}_{0,j}^\mu}{dt}\hbar S\frac{-J_j\uprm{H}}{2},
\end{align}
where $\mathcal{H}'\smrm{el}$ is the coefficient matrix of the Hamiltonian $H\smrm{el}(t)$ in $\tilde{c}_{\bm{k},j}$ representation.
By substituting the above equation into Eq.~\eqref{eq:COBoltzmann} and performing the Fourier transformation, we obtain the current density from the Fermi surface term $C_{\mathrm{O}(i,\mu)}^\lambda$ contribution as,
\begin{align}
  j_{\mathrm{O}}^\lambda (\omega) & =\sum_{j,\mu}i\omega \tau \sum_u \int\smrm{BZ} \frac{d\bm{k}}{(2\pi)^d} \mel{u}{j_{\bm{k}}^\lambda}{u} \mel{u}{\Pi_j \sigma_\mu}{u} \nonumber \\
 & \qquad  \times f_0'(\xi_{\bm{k},u}) \tilde{n}_{0,j}^\mu(\omega)\hbar S\frac{-J_j\uprm{H}}{2} \nonumber \\
 & =\sum_{j,\mu}i\hbar \omega C_{\mathrm{O}(i,\mu)}^\lambda \tilde{n}_{0,j}^\mu(\omega)\hbar S\frac{-J_j\uprm{H}}{2}.
\end{align}
Note that $\tilde{n}_{0,j}^\mu(\omega)\hbar S$ corresponds to $\tilde{S}_j^\mu$ in the Green's function formalism.

Next, we consider the limit of large Hund coupling $J\uprm{H}$ and approximate response current density $j_\mathrm{O}^\lambda$.
By doing so, we can extract the effect of the emergent electric field induced by the fluctuation of the localized spins.
Let us consider the Hamiltonian of the itinerant electrons subjected to the effective magnetic field from the time-dependent localized spins,
\begin{align}
  H\smrm{el}(t)= & -\sum_{\bm{k},j,\tau} [t_{j,\tau}e^{i\bm{k}\cdot \bm{r}_\tau}c_{\bm{k},j}^\dagger c_{\bm{k},j+\tau}+h.c.]\nonumber \\
 & \qquad -\sum_{\bm{k},j}\frac{J_j\uprm{H}\hbar S}{2}\bm{n}_j(t)\cdot c_{\bm{k},j}^\dagger \bm{\sigma} c_{\bm{k},j}\nonumber \\
  = & -\sum_{\bm{k},j,\tau} [t_{j,\tau}e^{i\bm{k}\cdot \bm{r}_\tau}\tilde{c}_{\bm{k},j}^\dagger(t) U_j^\dagger (t)U_j (t)\tilde{c}_{\bm{k},j+\tau}(t)+h.c.]\nonumber \\
 & \qquad -\sum_{\bm{k},j}\frac{J_j\uprm{H}\hbar S}{2}\tilde{c}_{\bm{k},j}^\dagger (t) \sigma_z \tilde{c}_{\bm{k},j}(t),
\end{align}
where $U_j(t)$ is the time-dependent extension of $U_j$ in Eq.~\eqref{eq:U},
\begin{align}
  U_j (t) & =\begin{pmatrix}
               e^{-i\phi_j(t)/2}\cos\frac{\theta_j(t)}{2} & e^{-i\phi_j(t)/2}\sin\frac{\theta_j(t)}{2} \\
               e^{i\phi_j(t)/2}\sin\frac{\theta_j(t)}{2} & -e^{i\phi_j(t)/2}\cos\frac{\theta_j(t)}{2}
             \end{pmatrix}\nonumber \\
 & =\left(
  \ket{\bm{n}_j(t)},\ket{-\bm{n}_j(t)}
  \right).
\end{align}
In the limit of large Hund coupling $t\ll \frac{J\uprm{H}\hbar S}{2}$, we can ignore the down (up) spin component in the rotated frame for the case of less (more) than the half filling.
Now, we consider the case of less than the half filling and project the Hamiltonian to the subspace where all the itinerant spins are up in the rotated frame,
\begin{align}
  H_{\mathrm{el}\uparrow}= & \sum_{\bm{k},j,\tau} -t_{j,\tau}e^{i\bm{k} \cdot \bm{r}_\tau}\tilde{c}_{\bm{k},j,\uparrow}^\dagger \braket{\bm n_j}{\bm n_{j+\tau}}\tilde{c}_{\bm{k},j+\tau,\uparrow}+h.c. \nonumber \\
 & \qquad -\sum_{\bm{k},j}\frac{J_j\uprm{H}\hbar S}{2}\tilde{c}_{\bm{k},j,\uparrow}^\dagger  \tilde{c}_{\bm{k},j,\uparrow}.\label{eq:HelUp}
\end{align}
In this subspace, we evaluate the Fermi surface terms and obtain the approximated response current density,
\begin{align}
  j_{\mathrm{O}}^\lambda & \simeq -\tau \sum_{u_\uparrow}\int\smrm{BZ} \frac{d\bm{k}}{(2\pi)^d} \mel{u_\uparrow}{j_{\bm{k}\uparrow}^\lambda}{u_\uparrow} \left. \frac{\partial \xi_{\bm{k},u\uparrow}}{\partial t}\right|_{t=0}f_0'(\xi_{\bm{k},u})\label{eq:approximatedCOBoltzmann}
\end{align}
with
\begin{align} 
  \left.\frac{\partial \xi_{\bm{k},u\uparrow}}{\partial t}\right|_{t=0} & =\left. \frac{\partial \mel{u_\uparrow}{\mathcal{H}_{\mathrm{el}\uparrow}}{u_\uparrow}}{\partial t}\right|_{t=0} \nonumber \\
 & =\mel{u_\uparrow}{\left. \frac{\partial \mathcal{H}_{\mathrm{el}\uparrow}}{\partial t}\right|_{t=0}}{u_\uparrow} \label{eq:timeDeivativeOfeigenE} \\
  j_{\bm{k}\uparrow}^\lambda & = -\frac{e}{\hbar}\partial_{\bm{k}_\lambda}\mathcal{H}_{\mathrm{el}\uparrow},
\end{align}
where $\mathcal{H}_{\mathrm{el}\uparrow}$ is the coefficient matrix of $H_{\mathrm{el}\uparrow}$ in $\tilde{c}_{\bm{k},j,\uparrow}(t)$ representation, and $\ket{u_\uparrow}$ is the eigen state of $\mathcal{H}_{\mathrm{el}\uparrow}$ in equilibrium ($t=0$).
In the second line of Eq.~\eqref{eq:timeDeivativeOfeigenE}, we use the Hellmann-Feynman theorem.

Finally, we discuss the physical meaning of the current density from the contribution of the Fermi surface terms.
From now on, we consider a 1D model whose hoppings have finite real values only between nearest neighbor sites for simplicity.
In this model, the Fermi surface term is interpreted as the emergent electric field in the limit of large Hund coupling.
For a 1D model whose hoppings have finite values only between nearest neighbor sites, we can employ the twisted parallel-transport gauge~\cite{Vanderbilt2018} to the Hamiltonian in equilibrium ($t=0$) in Eq.~\eqref{eq:HelUp},
\begin{align}
  H_{\mathrm{el}\uparrow}= & \sum_{k,j} \big[-t_{j,1}e^{i(k +\frac{e}{\hbar}A_0)a} |\braket{\bm n_j}{\bm n_{j+1}}|\tilde{c}_{\mathrm{twi}k,j}^\dagger\tilde{c}_{\mathrm{twi}k,j+1}\nonumber \\
 & \qquad \  +h.c. \big]                                                              
  -\sum_{k,j}\frac{J_j\uprm{H}\hbar S}{2}\tilde{c}_{\mathrm{twi}k,j}^\dagger  \tilde{c}_{\mathrm{twi}k,j},
\end{align}
with
\begin{align}
  \tilde{c}_{\mathrm{twi}k,j}\equiv & \begin{cases}
                                        \tilde{c}_{k,1,\uparrow} & j=1, \\
                                        e^{i\varphi_j}\tilde{c}_{k,j,\uparrow} & j\geq 2,
                                      \end{cases}\label{eq:twistedGaugeTrans} \\
  \varphi_{j}= & \sum_{l=1}^{j-1}(\mathrm{Im}\log \braket{\bm{n}_{l}(0)}{\bm{n}_{l+1}(0)}-\frac{e}{\hbar}A_0a) \\
  A_0\equiv & \frac{\hbar}{eN\smrm{site}a}\mathrm{Im}\log \prod_j \braket{\bm{n}_{j}(0)}{\bm{n}_{j+1}(0)},
\end{align}
where $a$ is the lattice constant.
Consequently, by denoting the coefficient matrix of $H_{\mathrm{el}\uparrow}$ in $\tilde{c}_{\mathrm{twi}k,j}$ representation as $\mathcal{H}_{\mathrm{twi}}$,
we can see the effective time-reversal symmetry,
\begin{align}
  \mathcal{H}_{\mathrm{twi}}^*(k-\frac{e}{\hbar}A_0) & =\mathcal{H}_{\mathrm{twi}}(-k-\frac{e}{\hbar}A_0)\label{eq:timeReversal} \\
  \xi_{k-\frac{e}{\hbar}A_0,u\uparrow} & =\xi_{-k-\frac{e}{\hbar}A_0,u\uparrow}\label{eq:timeReversalEigenE} \\
  \ket{u_{\mathrm{twi}}^*(-k-\frac{e}{\hbar}A_0)} & =\ket{u_{\mathrm{twi}} (k-\frac{e}{\hbar}A_0)}\label{eq:timeReversalEigenState},
\end{align}
where $\ket{u_{\mathrm{twi}}(k)}$ is the eigen state of $\mathcal{H}_{\mathrm{twi}}$.
This effective time-reversal symmetry appears in the system where we can perform the twisted parallel-transport gauge transformation Eq.~\eqref{eq:twistedGaugeTrans}.
In other words, this symmetry appears if no loop of bonds with finite hopping amplitudes
encloses emergent magnetic flux, i.e., $\sum_{j\in \text{loop}}\mathrm{Im}\log \braket{\bm{n}_j(0)}{\bm{n}_{j+\tau}(0)}=0$ for any loop.
By denoting the coefficient matrix of the gauge transformation Eq.~\eqref{eq:twistedGaugeTrans} as $U\smrm{twi}=\mathrm{diag}[1,e^{i\varphi_2},\dots,e^{i\varphi_{N\smrm{site}}}]$, we obtain the relation between $\ket{u_\uparrow}$ and $\ket{u_{\mathrm{twi}}}$ as $\ket{u_{\mathrm{twi}}}=U\smrm{twi}\ket{u_\uparrow}$.
Using these expressions, we can rewrite the approximated response current density Eq.~\eqref{eq:approximatedCOBoltzmann} as,
\begin{align}
  j_{\mathrm{O}}\simeq & -\tau \sum_{u_\uparrow}\int\smrm{BZ}\frac{dk}{2\pi}\mel{u_\uparrow}{j_k}{u_\uparrow}f_0'(\xi_{k\uparrow})\mel{u_\uparrow}{\left. \frac{\partial \mathcal{H}\smrm{el}}{\partial t}\right|_{t=0}}{u_\uparrow}\nonumber \\
  = & -\frac{e\tau}{2\pi \hbar }\sum_{u\smrm{twi}}\sum_{k_i \in \mathrm{RM}}\mel{u\smrm{twi}}{U\smrm{twi}\left.\frac{\partial \mathcal{H}_{\mathrm{el}\uparrow}}{\partial t}\right|_{t=0}U\smrm{twi}^\dagger }{u\smrm{twi}}\nonumber \\
 & +\frac{e\tau}{2\pi \hbar }\sum_{u\smrm{twi}}\sum_{k_i \in \mathrm{LM}}\mel{u\smrm{twi}}{U\smrm{twi}\left.\frac{\partial \mathcal{H}_{\mathrm{el}\uparrow}}{\partial t}\right|_{t=0}U\smrm{twi}^\dagger }{u\smrm{twi}},
\end{align}
where RM (LM) is the set of the Fermi wave numbers labeling right (left) mover electrons, defined by the condition $\mel{u_\uparrow}{j_k}{u_\uparrow}>0$ ($\mel{u_\uparrow}{j_k}{u_\uparrow}<0$).
Consequently, using Eqs.~\eqref{eq:timeReversalEigenE} and \eqref{eq:timeReversalEigenState}, we obtain,
\begin{align}
  j_{\mathrm{O}}\simeq & -\frac{e\tau}{2\pi \hbar }\sum_{u\smrm{twi}}\sum_{k_i \in \mathrm{RM}}\mel{u\smrm{twi}}{\mathcal{A}}{u\smrm{twi}}, \label{eq:onlyMathcalA}
\end{align}
with
\begin{align}
  \mathcal{A}(k_i)= & U\smrm{twi}\left.\frac{\partial \mathcal{H}_{\mathrm{el}\uparrow}}{\partial t}\right|_{t=0,k=k_i}U\smrm{twi}^\dagger \nonumber \\
 & -\left[U\smrm{twi}\left.\frac{\partial \mathcal{H}_{\mathrm{el}\uparrow}}{\partial t}\right|_{t=0,k=-k_i-2\frac{e}{\hbar }A_0}U\smrm{twi}^\dagger\right]^*.
\end{align}
Note that if $k_i$ is the Fermi wave number and labels right (left) mover electron, $-k_i-2\frac{e}{\hbar}A_0$ is the Fermi wave number and labels left (right) mover electron because of Eq.~\eqref{eq:timeReversalEigenE}.

Now that we have shown that $\mathcal{A}$ makes the Fermi surface terms finite, let us discuss the physical meaning of response current density stemming from $\mathcal{A}$.
We will show that
the finite value of $\mathcal{A}$ corresponds to the finite value of the emergent electric field.
$\mathcal{A}$ is given as,
\begin{gather}
  \mathcal{A}_{m,n}=  \begin{cases}
                         -2t_{m,1}(k)i\partial_t\mathrm{Im}\log \braket{\bm{n}_m}{\bm{n}_{m+1}} & n=m+1, \\
                         -2t_{m-1,1}^*(k)i\partial_t\mathrm{Im}\log \braket{\bm{n}_m}{\bm{n}_{m-1}} & n=m-1, \\
                         0 & \text{otherwise},
                       \end{cases}\label{eq:AsymmetricComponent} \\
  t_{m,1}(k)=  t_{m,1}e^{i(k +\frac{e}{\hbar}A_0)a}|\braket{\bm{n}_m}{\bm{n}_{m+1}}|.
\end{gather}
Here, we go back to Eq.~\eqref{eq:HelUp} and focus on the factor $\braket{\bm{n}_j}{\bm{n}_{j+1}}$ in Eq.~\eqref{eq:HelUp}.
It can be rewritten as
\begin{gather}
  \braket{\bm{n}_j}{\bm{n}_{j+1}}=  e^{\varphi(t)}, \\
  \varphi(t)=  \mathrm{Re}\log\braket{\bm{n}_j}{\bm{n}_{j+1}}+i\mathrm{Im}\log\braket{\bm{n}_j}{\bm{n}_{j+1}}.\label{eq:varphiT}
\end{gather}
Here, $\mathrm{Im}\log\braket{\bm{n}_j}{\bm{n}_{j+1}}$ is the Berry connection in the real space.
As can be seen from Eq.~\eqref{eq:AsymmetricComponent}, the time derivative of the second term on the right hand side of Eq.~\eqref{eq:varphiT} produces $\mathcal{A}$ and the time derivative of the first term cancels out in Eq.~\eqref{eq:onlyMathcalA}.
Therefore, in the following, we discuss the physical meaning of $e^{i\mathrm{Im}\log\braket{\bm{n}_j}{\bm{n}_{j+1}}}$ and $\partial_t\mathrm{Im}\log\braket{\bm{n}_j}{\bm{n}_{j+1}}$.

We consider $e^{i\mathrm{Im}\log\braket{\bm{n}_j}{\bm{n}_{j+1}}}$ as the effective Peierls phases that cause the effective gauge fields and drive the Bloch electron at $k$.
Specifically, we define the emergent vector potential between site $j$ and $j+1$ as $A_{j,j+1}\uprm{em}=\frac{\hbar }{ea}\mathrm{Im} \log  \braket{\bm n_j}{\bm n_{j+1}}$,
and the emergent electric potential at site $j$ as $V_j=\frac{1}{e}\mel{\bm{n}_j}{i\hbar \partial_t}{\bm{n}_j}$.
By defining these emergent electromagnetic potentials, the emergent electric field experienced by Bloch electron at $k$ is given by
\begin{align}
 & \frac{1}{N\smrm{site}a}\sum_{j}^{N\smrm{site}} [V_{j}-V_{j+1}+(\bm{r}_j-\bm{r}_{j+1})\cdot (-\partial_t A_{j,j+1})] \nonumber \\
  = & \frac{\hbar}{eN\smrm{site}a}\partial_t\mathrm{Im} \log  \prod_j \braket{\bm n_j}{\bm n_{j+1}},\label{eq:emergentE}
\end{align}
where $N\smrm{site}$ is the number of the sites in a magnetic unit cell.
This equation shows that the emergent electric field is generated by the time derivative of the Berry phase in real space.
In this picture, $C_{\mathrm{O}(i,\mu)}^\lambda$ can be interpreted as energy transfer between the itinerant electrons and the localized spins via Berry phase in real space.  
Figure~\ref{fig:schematic}(b) is the schematic picture of the time variation of the Berry phase induced by the spin configuration, the emergent electric field.
This can be considered by analogy with the time derivative of the magnetic flux surrounded by a coil generating the inductance~\cite{Barnes2007}.
Furthermore, Eq.~\eqref{eq:emergentE} becomes the conventional emergent electric field in the continuum limit
\begin{align}
  \frac{\hbar}{eN\smrm{site}a}\partial_t\mathrm{Im} \log  \prod_j \braket{\bm n_j}{\bm n_{j+1}}
  \to
  \frac{1}{N\smrm{site}a}\int_{\bm{r}_1}^{\bm{r}_n} \bm{E}\smrm{em}\cdot d\bm{r},\label{eq:emergentEtoContinuum}
\end{align}
where $\bm{E}\smrm{em}=\frac{\hbar}{2e}\bm{n}\cdot (\partial_x \bm{n} \times \partial_t \bm{n})=\frac{\hbar}{2e}\sin \theta (\dot{\theta}\partial_x \phi -\dot{\phi}\partial_x \theta)$ is the emergent electric field in continuum systems~\cite{Volovik1987}.
For details of derivation, see Appendix~\ref{app:emergentE}.
In this expression, one can see that the emergent electric field drives the spin current.
For more than half filling case,
the opposite-direction emergent electric field arises because $ H_{\mathrm{el}\downarrow}(\{\bm{n}_j\}) =H_{\mathrm{el}\uparrow}(\{-\bm{n}_j\})$ and  $\braket{\bm{n}_j}{\bm{n}_{j+\tau}}=\braket{-\bm{n}_j}{-\bm{n}_{j+\tau}}^*$.
Therefore, the emergent electric field drives the spin current in the rotated frame.
In the above discussion, we consider 1D models whose hoppings have finite real values only between nearest neighbor sites.
However, the discussion can be extended to 1D models in which further-neighboring sites have finite complex hopping amplitudes as far as no loop of bonds has finite magnetic fluxes.
This is because the twisted parallel-transport gauge Eq.~\eqref{eq:twistedGaugeTrans} can also be employed in this situation.

\subsubsection{Interband effect terms $C_{\mathrm{I}(i,\mu)}^\lambda$}
In this section, we discuss the interband effect terms $C_{\mathrm{I}(i,\mu)}^\lambda$.
This term, in contrast, corresponds to the current response induced by the time variation of the polarization, $j_{\mathrm{I}}^\lambda (t)\equiv \dot{P}^\lambda$,
which can be expressed by the integral of Berry curvature in $k-t$ space~\cite{Resta1994}.
Specifically, the time derivative of the polarization density is given by
\begin{align}
  \dot{P}^\lambda=-2\hbar \mathrm{Im}\sum_{u\neq v}\int \frac{d\bm{k}}{(2\pi)^d}\frac{\mel{u}{{j}_{\bm{k}}^\lambda}{v}\mel{v}{\dot{\mathcal{H}}\smrm{c}}{u}}{(\xi_{\bm{k},u}-\xi_{\bm{k},v})^2}f(\xi_{\bm{k},u}),
\end{align}
where $\mathcal{H}\smrm{c}$ is the coefficient matrix of $H\smrm{c}(t)$ in Eq.~\eqref{eq:HcTimeDependent} in $\tilde{c}_{\bm{k},j}$ representation.
We can derive $C_{\mathrm{I}(j,\mu)}^\lambda$ in Eq.~\eqref{eq:CI} by substituting $\mathcal{H}\smrm{c}$ and performing the Fourier transformation in time as,
\begin{align}
  j_\mathrm{I}^\lambda(\omega) = & \sum_{j,\mu}i\hbar\omega 2 \mathrm{Im}\sum_{u\neq v}\int \frac{d\bm{k}}{(2\pi)^d}\frac{\mel{u}{{j}_{\bm{k}}^\lambda}{v}\mel{v}{\Pi_j \sigma_\mu}{u}}{(\xi_{\bm{k},u}-\xi_{\bm{k},v})^2}\nonumber \\
 & \qquad \times f(\xi_{\bm{k},u})\tilde{n}_{0,j}^\mu(\omega)\hbar S\frac{-J_j\uprm{H}}{2},\nonumber \\
  = & i\hbar\omega \sum_{j,\mu} C_{\mathrm{I}(j,\mu)}^\lambda \tilde{n}_{0,j}^\mu(\omega)\hbar S\frac{-J_j\uprm{H}}{2}.
\end{align}
Note that $\tilde{n}_{0,j}^\mu(\omega)\hbar S$ corresponds to $\tilde{S}_j^\mu$ in the Green's function formalism.

\subsection{Green's function $D$ and conductivity $\Sigma$}\label{subsec:GreenFunction}
In this section, we evaluate the conductivity $\Sigma^{\lambda \rho}(\omega)$ by using the Green's function $D(\omega)$ and obtain the impedance $Z^{\lambda \rho}(\omega )$ inductance $L^{\lambda \rho}$ from the conductivity.

First, we derive the Green's function $D(\omega)$ in the bilinear form whose indices are $(i,\mu)$.
In other words, we express $H\smrm{m}$ in Eq.~\eqref{eq:Hmkspace} in the bilinear form of $\tilde{S}_{i}^\mu$ where $\mu=x,y$ as follows:
\begin{align}
  H\smrm{m}= & -\sum_{\substack{j,\tau, \\\mu=x,y,\\\nu=x,y}} \frac{1}{2}\tilde{J}_{j,\tau}^{\mu\nu}(\tilde{S}_{j}^\mu \tilde{S}_{j+\tau}^\nu+\tilde{S}_{j+\tau}^\nu\tilde{S}_{j}^\mu )        \nonumber \\
 & +\sum_{j,\tau,\mu=x,y} \frac{1}{2}(\tilde{J}_{j,\tau}^{zz}+\tilde{J}_{j-\tau,\tau}^{zz})\tilde{S}_{j}^{\mu2} \nonumber \\
 & +\sum_{j,\mu=x,y} \frac{1}{2}\frac{\tilde{B}_j^z}{\hbar S}\tilde{S}_{j}^{\mu2}+const. \label{eq:HmBilinear}
\end{align}
Here, we omit the label $\bm{k}(=0)$ and approximate the spin operators using the Holstein-Primakoff transformation.
Namely, we perform the Holstein-Primakoff transformation to Eq.~\eqref{eq:Hmkspace} and express the magnon Hamiltonian in the bilinear form of $\tilde{S}_j^{\mu=x,y}$.
The constant terms in Eq.~\eqref{eq:HmBilinear} include c-number terms proportional to $S$ or $S^2$.
For detailed derivation, see the derivation above Eq.~\eqref{eqApp:HmMag} in Appendix~\ref{app:magnonFormalism}.
By denoting $\mathcal{H}\smrm{m}(\bm{k}=0)$ as the coefficient matrix of $H\smrm{m}$ in $\tilde{S}_{j}^{\mu=x,y}$ representation in Eq.~\eqref{eq:HmBilinear}, we obtain the Green's function $D(\omega)$ as follows:
\begin{align}
  D(\omega)=(-\sigma_y\hbar \omega +i\alpha \hbar \omega -\mathcal{H}\smrm{m}(\bm{k}=0))^{-1},\label{eq:Green}
\end{align}
where $\alpha$ is the Gilbert damping.
Here, we adopt the Rayleigh dissipation function $\sum_i \frac{\alpha}{2\hbar S}|\dot{\bm S_i}|^2$ to introduce the Gilbert damping term~\cite{Gilbert2004}.
Note that the expression of the Green's function in $\tilde{S}_{j}^\mu$ representation is different from the one in the magnon representation that is given as $(\sigma_z\hbar \omega +i\alpha \hbar \omega-\mathcal{H}\uprm{mag}(\bm{k}=0))^{-1}$.

By substituting Eqs.~\eqref{eq:sigmaIMuWithC} and \eqref{eq:chiJNuWithC} into Eq.~\eqref{eq:Sigma}, we obtain the conductivity mediated by magnon $\Sigma^{\lambda \rho}(\omega)$,
\begin{align}
  \Sigma^{\lambda \rho} (\omega)= & i\hbar ^2 \omega v \sum_{i,j,\mu,\nu} (C_{\mathrm{O}(i,\mu)}^\lambda+C_{\mathrm{I}(i,\mu)}^\lambda)\frac{-J_i\uprm{H}}{2}\nonumber \\
 & \qquad \times D_{(i,\mu),(j,\nu)}(\omega)\frac{-J_j\uprm{H}}{2}(-C_{\mathrm{O}(j,\nu)}^\rho+C_{\mathrm{I}(j,\nu)}^\rho).
  \label{eq:SigmaWithC}
\end{align}
Subsequently, by the definition of the impedance $Z^{\lambda \rho} (\omega)\equiv (l/A)(\sigma \smrm{dc}^{\lambda \rho}+\Sigma^{\lambda \rho}  (\omega))^{-1}$ and the inductance
\begin{align}
  L^{\lambda \rho} & \equiv -\mathrm{Im}[\partial_\omega Z (\omega)]_{\omega=0}\nonumber \\
 & =\frac{l}{A}\frac{\mathrm{Im}[\partial_\omega \Sigma^{\lambda \rho}] _{\omega=0}}{(\sigma\smrm{dc}^{\lambda \rho})^2} \label{eq:inductance}
\end{align}
with $l$ ($A$) being a sample length (cross section), we obtain the inductance
\begin{align}
  L^{\lambda \rho} = & \frac{\hbar ^2l v }{A(\sigma\smrm{dc}^{\lambda \rho})^2}  \sum_{i,j,\mu,\nu} (C_{\mathrm{O}(i,\mu)}^\lambda+C_{\mathrm{I}(i,\mu)}^\lambda)\frac{-J_i\uprm{H}}{2}\nonumber \\
 & \qquad \times D_{(i,\mu),(j,\nu)}(0)\frac{-J_j\uprm{H}}{2}(-C_{\mathrm{O}(j,\nu)}^\rho+C_{\mathrm{I}(j,\nu)}^\rho).
\end{align}

\begin{figure}
  \includegraphics[width=\linewidth]{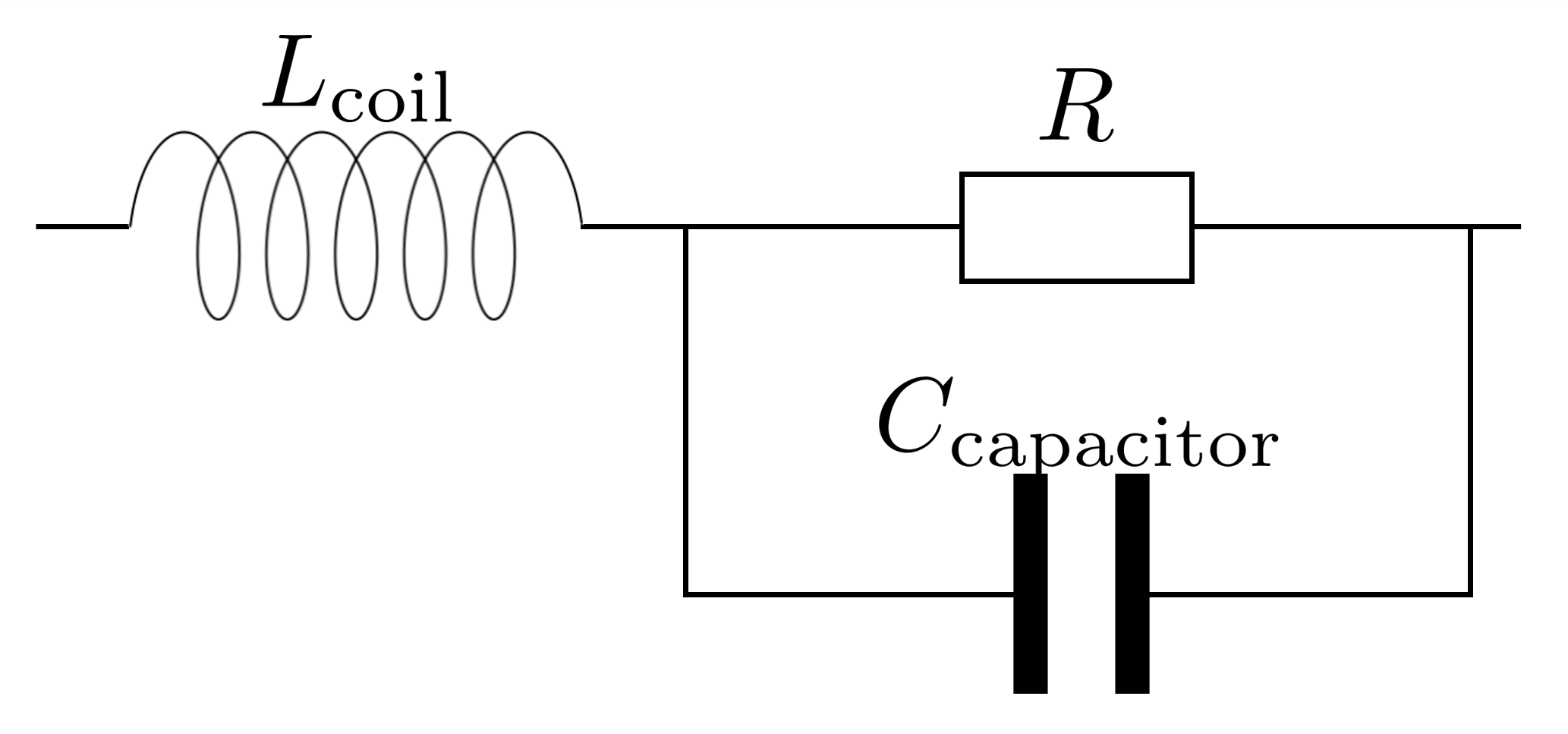}
  \caption{The circuit diagram of the classical circuit describing the inductance $L^{\lambda \lambda}$ in Eq.~\eqref{eq:decomposedL}. The circuit consists of a resistor and capacitor connected in parallel, which are then connected in series with the coil.}
  \label{fig:classicalCircuit}
\end{figure}

Now, we focus on the longitudinal inductance $L^{\lambda\lambda}$
and discuss the contribution to $L^{\lambda\lambda}$ from the Fermi surface terms and the interband effect terms.
The Green's function $D_{(i,\mu),(j,\nu)}(0)$ is a symmetric matrix because $H\smrm{m}$ is a symmetric matrix.
(For details, see the discussion below Eq.~\eqref{eqApp:HmMag} in Appendix~\ref{app:magnonFormalism}.)
Therefore, the inductance $L^{\lambda\lambda}$ is decomposed as following,
\begin{align}
  &L^{\lambda\lambda}=\nonumber \\
  & \frac{\hbar ^2l v }{A(\sigma\smrm{dc}^{\lambda \lambda})^2}\sum_{i,j,\mu,\nu} C_{\mathrm{O}(i,\mu)}^\lambda\frac{-J_i\uprm{H}}{2}(-D_{(i,\mu),(j,\nu)}(0))\frac{-J_j\uprm{H}}{2}C_{\mathrm{O}(j,\nu)}^\lambda\nonumber \\
 & -\frac{\hbar ^2l v }{A(\sigma\smrm{dc}^{\lambda \lambda})^2}  \sum_{i,j,\mu,\nu} C_{\mathrm{I}(i,\mu)}^\lambda\frac{-J_i\uprm{H}}{2}(-D_{(i,\mu),(j,\nu)}(0))\frac{-J_j\uprm{H}}{2}C_{\mathrm{I}(j,\nu)}^\lambda. \label{eq:decomposedL}
\end{align}
Here, the eigenenergies of $H\smrm{m}$ must be positive because $\tilde{S}_{i,\mu}$ are perturbations from the ground state and there is a finite gap to the first excited state.
Therefore, $H\smrm{m}$ is positive definitive because of Sylvester's law of inertia~\cite{Colpa1978} and consequently $-D(0)$ is positive definitive.
Therefore, the first term and the second term in Eq.~\eqref{eq:decomposedL} are always positive and negative respectively.
The first term is the contribution from the Fermi surface terms $C_{\mathrm{O}(i,\mu)}^\lambda$ which include the effect of the Ohmic dissipation driven by the emergent electric field.
The emergent electric field corresponds to the time variation of Berry phase induced by the dynamical spin configuration as shown in Fig.\ref{fig:schematic}(b).
Here, the Berry phase plays a role of magnetic flux inside a coil.
In contrast, the second term is the contribution from the interband effect terms $C_{\mathrm{I}(i,\mu)}^\lambda$ which corresponds to the polarization induced by AC field serving as a capacitor.
Therefore, $L^{\lambda\lambda}$ has the two components: the positive (inductive) contribution including the effect of the emergent electric field, and the negative (capacitive) contribution due to the electric polarization induced by the spin dynamics.
This situation can be interpreted by a classical circuit with a coil $L\smrm{coil}$, a capacitor $C\smrm{capacitor}$ and a resistance $R$, as depicted in Fig.~\ref{fig:classicalCircuit}.
The impedance of the circuit element in Fig.~\ref{fig:classicalCircuit} is given as
\begin{align}
    Z&=-i\omega L\smrm{coil}+\left(\frac{1}{R}-i\omega C\smrm{capacitor}\right)^{-1}\nonumber \\
    &=R-i\omega (L\smrm{coil}-C\smrm{capacitor}R^2) +O(\omega^2),
\end{align}
where $L\smrm{coil}$ corresponds to the positive contribution in $L^{\lambda\lambda}$ and $C\smrm{capacitor}$ corresponds to the negative contribution in $L^{\lambda\lambda}$.

\section{Numerical results}\label{sec:InductanceAndImpedance}

In this section, we present the numerical results for the inductances, the impedances, the spin dynamics, and the Q-values for various quasi 1D models shown in Fig.~\ref{fig:models}.
Specifically, we apply the present method to four magnets: helical states, fan states, 1D kagome models and canted-antiferromagnetic (AFM) states with the Rice-Mele model.
We adopt the helical state as a reference, the fan state as a case that does not induce time variation of a net magnetization along the q-vector direction, the 1D Kagome model, which does not have the Dzyaloshinskii-Moriya (DM) interaction, as a SOC free model in spin Hamiltonian, and the canted-AFM as a model which does not generate an emergent electric field.
For considering quasi 1D models in the present method, we adopt $d=1$ in Eqs.~\eqref{eq:CO} and \eqref{eq:CI} and compute the conductivities by dividing $\sigma\smrm{dc}$ and $\Sigma(\omega)$ in 1D model with $a^2$ ($a$: lattice constant).

\subsection{1D models}\label{subsec:1Dmodels}
\begin{figure}
  \includegraphics[width=\linewidth]{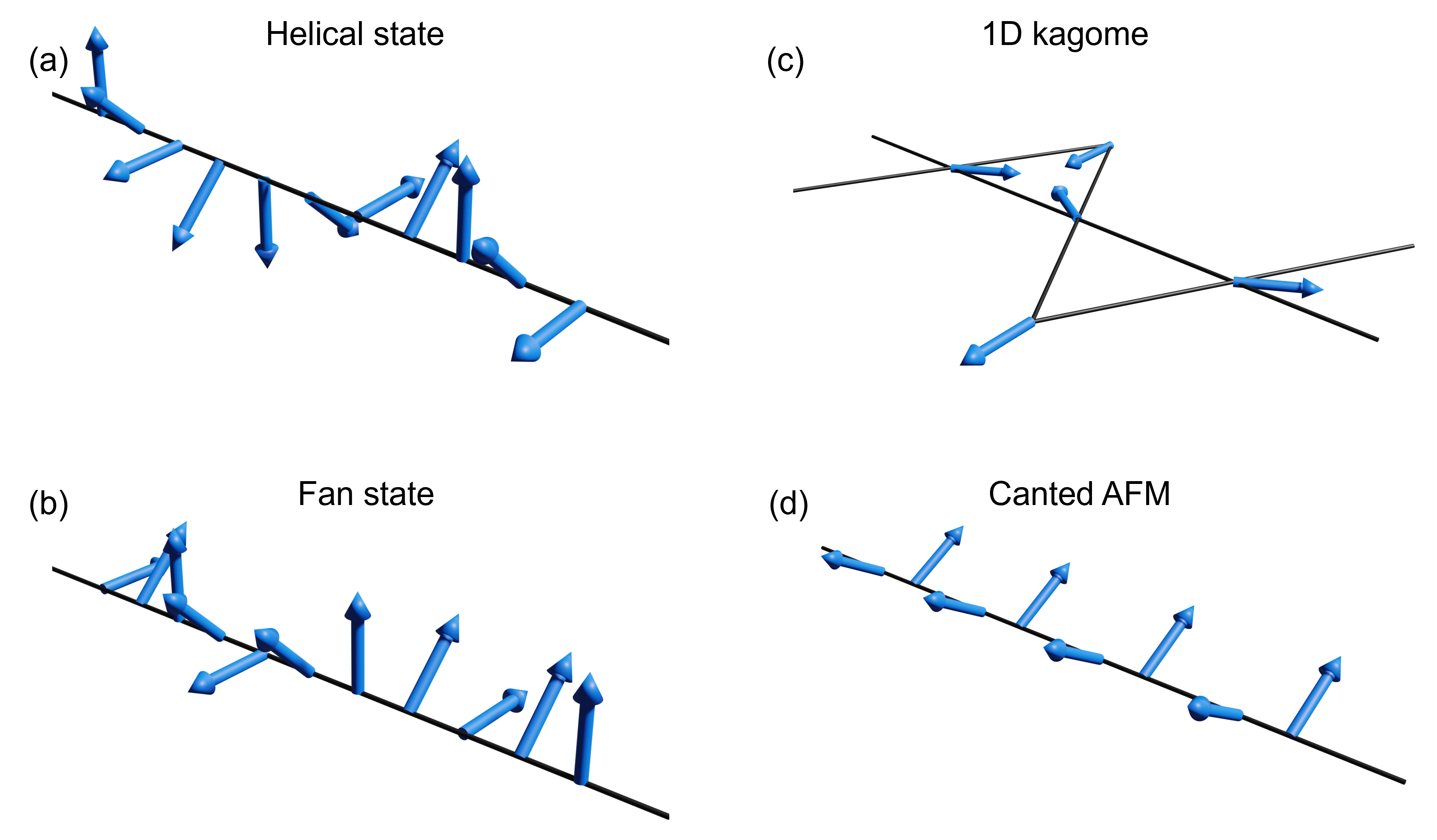}
  \caption{Schematic pictures of (a) the helical state, (b) the fan state, (c) the 1D kagome model, and (d) the canted-antiferromagnetic state with the Rice-Mele model.}
  \label{fig:models}
\end{figure}

We perform the numerical calculations for the quasi 1D models shown in Fig.~\ref{fig:models}.
These systems are described by the Hamiltonians
\begin{align}
  H=H\smrm{itinerant}+H\smrm{Hund}+H\smrm{m}.
\end{align}
Here, we adopt the spacially uniform Hund coupling
\begin{align}
  H\smrm{Hund}=-\sum_{i}\frac{J\uprm{H}}{2}\bm{S}_i \cdot c_i^\dagger \bm{\sigma} c_i,
\end{align}
for all the models,
while $H\smrm{itinerant}$ and $H\smrm{m}$ are different for each model.
In this section, we present $H\smrm{itinerant}$ and $H\smrm{m}$ and the spin configurations for four different magnetic states below.

(i) Helical state: the Hamiltonians of the helical state are given as follows,
\begin{gather}
  H\smrm{itinerant}\uprm{helical}  =\sum_{i}[-t_0c_{i}^\dagger c_{i+1}+h.c.], \\
  H\smrm{m}\uprm{helical}          =\sum_{i}[-J\smrm{ex}\bm{S}_i\cdot \bm{S}_{i+1} -D(\bm{S}_i\times \bm{S}_{i+1})_z -p\tilde{S}_i^{z2}],
\end{gather}
where $J\smrm{ex}$ is the exchange interaction, $D$ is the Dzyaloshinskii-Moriya interaction, $t_0$ is the hopping amplitude, and $p$ is the pinning potential.
The spin configuration of the helical state (i.e., the spin configuration of the ground state of $H\smrm{m}\uprm{helical}$) is described (Fig.~\ref{fig:models}(a)) as $(\theta_j,\phi_j)=(\pi/2,2\pi j/N\smrm{site})$.
Here, we choose the $z$ axis along the 1D chain.
The condition that the spin configuration of the ground state of $H\smrm{m}\uprm{helical}$ becomes this expression is given by $D=J\smrm{ex}\tan (2\pi/N\smrm{site})$.
By performing the orthogonal transformation Eq.~\eqref{eq:T}, we obtain $\tilde{J}_{j,\tau}$ as,
\begin{align}
  \begin{array}{c}
    \tilde{J}_{j,1}^{xx}=J\smrm{ex}, \\ \tilde{J}_{j,1}^{yy}=\tilde{J}_{j,1}^{zz}=\sqrt{J\smrm{ex}^2+D^2},\
    \tilde{J}_{j,0}^{zz}=p.
  \end{array} \label{eq:tildeJhelical}
\end{align}

(ii) Fan state: the Hamiltonians of the fan state are given as follows,
\begin{gather}
  H\smrm{itinerant}\uprm{fan}  =\sum_{i}[-t_0c_{i}^\dagger c_{i+1}+h.c.], \\
  H\smrm{m}\uprm{fan}          =\sum_{i}[-J\smrm{ex}\bm{S}_i\cdot \bm{S}_{i+1} -D\eta_i\uprm{fan}(\bm{S}_i\times \bm{S}_{i+1})_z -p\tilde{S}_i^{z2}], \label{eq:fanHm}\\
  \eta_i\uprm{fan}             =\begin{cases}
    +1 & i\leq N\smrm{site}/2 \mod N\smrm{site} \\
    -1 & i> N\smrm{site}/2 \mod N\smrm{site}
  \end{cases}.
\end{gather}
The spin configuration of the fan state is described (Fig.~\ref{fig:models}(b)) as $(\theta_j,\phi_j)=(\pi/2,\pi/2-|\pi-2\pi j/N\smrm{site}|)$.
Here, we choose the $z$ axis along the 1D chain.
The condition that the spin configuration of the ground state of $H\smrm{m}\uprm{fan}$ becomes this expression is given by $D=J\smrm{ex}\tan (2\pi/N\smrm{site})$.
By performing the orthogonal transformation Eq.~\eqref{eq:T}, we obtain $\tilde{J}_{j,\tau}$ as,
\begin{align}
  \begin{array}{c}
    \tilde{J}_{j,1}^{xx}=J\smrm{ex}, \\ \tilde{J}_{j,1}^{yy}=\tilde{J}_{j,1}^{zz}=\sqrt{J\smrm{ex}^2+D^2},\
    \tilde{J}_{j,0}^{zz}=p.
  \end{array}
\end{align}

(iii) 1D kagome model: the Hamiltonians of the 1D kagome model are given as follows,
\begin{gather}
  H\smrm{itinerant}\uprm{kagome}  =\sum_{i}[-t_0c_{i}^\dagger c_{i+1}-t_0c_{2i-1}^\dagger c_{2i+1} +h.c.], \\
  H\smrm{m}\uprm{kagome}          =\sum_{i}[J\smrm{ex}\bm{S}_i\cdot \bm{S}_{i+1} +J\smrm{ex}\bm{S}_{2i-1}\cdot \bm{S}_{2i+1} -p\tilde{S}_i^{z2}].
\end{gather}
The spin configuration of the 1D kagome model is described (Fig.~\ref{fig:models}(c)) as,
\begin{align}
  \theta_{j=1,2,3,4} & =\pi/2, \\
  (\phi_{1},\phi_{2},\phi_{3},\phi_{4}) & =(\pi / 3, 3\pi  / 2, 2\pi  / 3, 3\pi /2 ).
\end{align}
Here, we choose the $y$ axis along the 1D chain.
By performing the orthogonal transformation Eq.~\eqref{eq:T}, we obtain $\tilde{J}_{j,\tau}$ as,
\begin{gather}
  \tilde{J}_{j,1}^{xx}=\tilde{J}_{j,1}^{yy}=-\frac{1}{2}J\smrm{ex},\\
  \tilde{J}_{j,1}^{zz}=J\smrm{ex},\\
  \begin{cases}
    \tilde{J}_{j,1}^{xy}=-\tilde{J}_{j,1}^{yx}=-\frac{\sqrt{3}}{2}J\smrm{ex} & \cdots j=1,2 \\
    \tilde{J}_{j,1}^{xy}=-\tilde{J}_{j,1}^{yx}=\frac{\sqrt{3}}{2}J\smrm{ex} & \cdots j=3,4
  \end{cases},\\
  \tilde{J}_{j,2}^{zz}=J\smrm{ex},\\
  \tilde{J}_{1,2}^{xx}=\tilde{J}_{1,2}^{yy}=
  \tilde{J}_{3,2}^{xx}=\tilde{J}_{3,2}^{yy}=-\frac{1}{2}J\smrm{ex},\\
  \tilde{J}_{1,2}^{xy}=-\tilde{J}_{1,2}^{yx}=
  -\tilde{J}_{3,2}^{xy}=+\tilde{J}_{3,2}^{yx}=\frac{\sqrt{3}}{2}J\smrm{ex},\\
  \tilde{J}_{j,0}^{zz}=p.
\end{gather}

(iv) Canted-antiferromagnetic state with the Rice-Mele model: the Hamiltonians of the canted-AFM state are given as follows,
\begin{gather}
  H\smrm{itinerant}\uprm{cant}  =\sum_{i}[-(t_0+\delta t)c_{2i}^\dagger c_{2i+1}-(t_0-\delta t)c_{2i+1}^\dagger c_{2i+2}\nonumber\\
    \hspace{4em} +h.c.]+\sum_i V (c_{2i}^\dagger  c_{2i}-c_{2i+1}^\dagger  c_{2i+1}), \\
  H\smrm{m}\uprm{cant}          =\sum_{i}[J\smrm{ex}\bm{S}_i\cdot \bm{S}_{i+1} -D\eta_i\uprm{cant}(\bm{S}_i\times \bm{S}_{i+1})_z -p\tilde{S}_i^{z2} ],\\
  \eta_i\uprm{cant}             =\begin{cases}
    +1 & i=0 \mod 2 \\
    -1 & i=1 \mod 2
  \end{cases}.
\end{gather}
The spin configuration of the canted-AFM state is described (Fig.~\ref{fig:models}(d)) as,
\begin{align}
  \theta_j & =\pi/2, \\
  (\phi_{2j+1},\phi_{2j}) & =(-\pi/2+\phi\smrm{cant}, \pi/2-\phi\smrm{cant} ),
\end{align}
where $\phi\smrm{cant}$ is the canting angle from the antiferromagnetic state.
Here, we choose the $z$ axis along the 1D chain.
The condition that the spin configuration of the ground state of $H\smrm{m}$ becomes the spin configuration represented by the above equation is $D=-J\smrm{ex}\tan (2\phi\smrm{cant})$.
By performing the orthogonal transformation Eq.~\eqref{eq:T}, we obtain $\tilde{J}_{j,\tau}$ as,
\begin{align}
  \begin{array}{c}
    \tilde{J}_{j,1}^{xx}=-J\smrm{ex}, \\ \tilde{J}_{j,1}^{yy}=\tilde{J}_{j,1}^{zz}=\sqrt{J\smrm{ex}^2+D^2},\
    \tilde{J}_{j,0}^{zz}=p.
  \end{array}
\end{align}

\begin{figure*}
  \includegraphics[width=\linewidth]{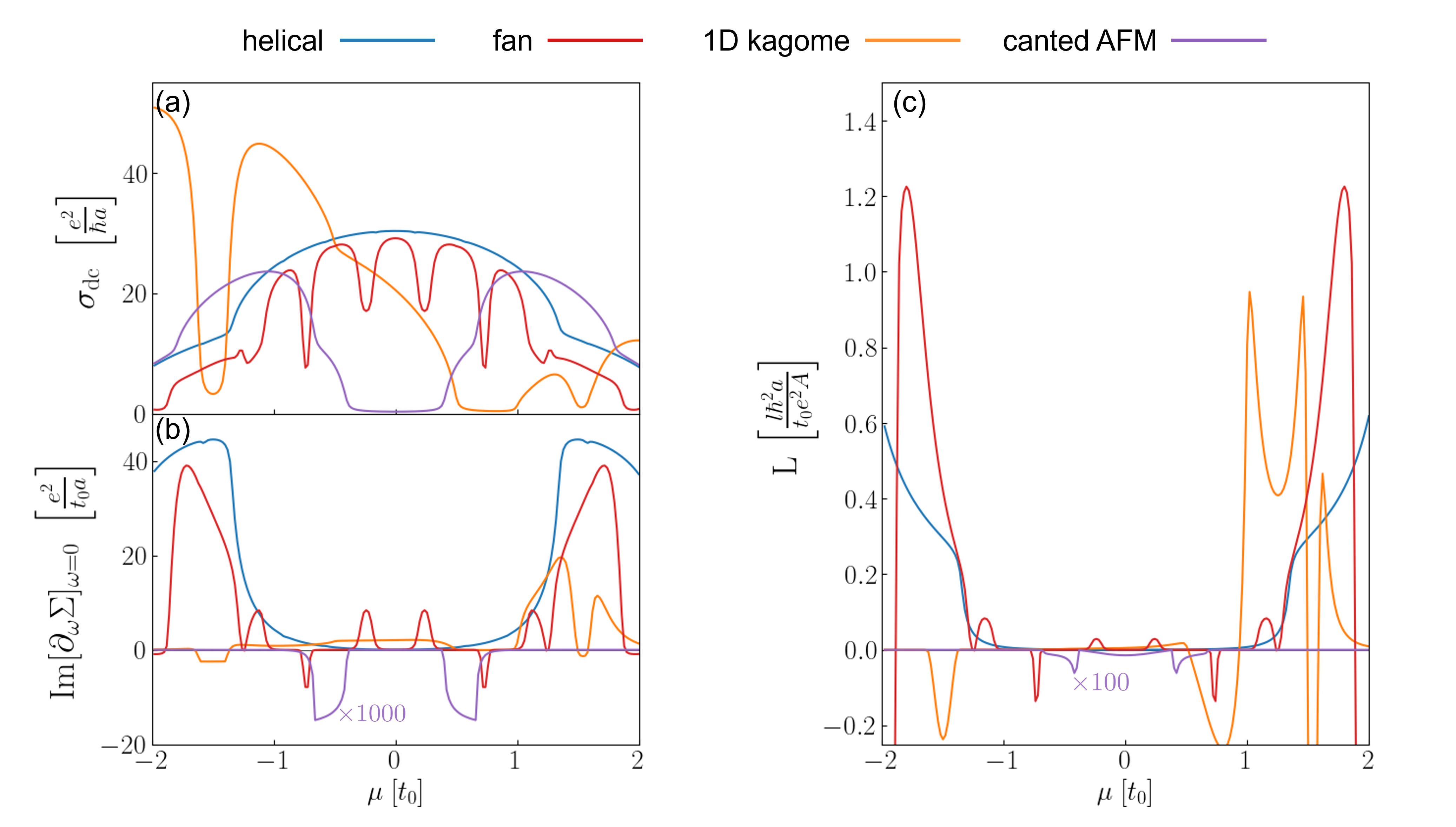}
  \caption{(a) DC conductivity $\sigma\smrm{dc}$, (b) the imaginary part of the frequency derivative of the conductivity mediated by magnon $\mathrm{Im}[\partial_\omega \Sigma] _{\omega=0}$ and (c) the inductance $L$ as a function of the chemical potential $\mu$ for the helical state (blue), the fan state (red), the 1D kagome models (orange), and the canted-antiferromagnetic state with the Rice-Mele model (purple).
    We adopt the parameter set: $J\smrm{ex}=0.04t_0,J\uprm{H}=0.5t_0,p=0.04t_0,\tau=25\hbar /t_0,\delta t =0.1t_0,V=0.1t_0,\phi\smrm{cant}=\pi/6$ and $N\smrm{site}=8$ for the helical state and the fan state.}
  \label{fig:Inductance}
\end{figure*}

\subsection{Inductance}\label{subsec:Inductance}
Figure~\ref{fig:Inductance}(a) shows the DC conductivity $\sigma\smrm{dc}$.
The helical state has no energy gap and the other models have energy gaps in the region of $-2t_0<E<2t_0$.
Correspondingly, there are some dip structures in the DC conductivity due to the gaps or the decreases of density of states except for the cases of the helical state.

Figure~\ref{fig:Inductance}(b) shows the imaginary part of the frequency derivative of the conductivity mediated by magnon, which is given as
\begin{align}
 & \mathrm{Im}[\partial_\omega \Sigma] _{\omega=0}\nonumber \\
  = & \hbar^2 a\sum_{i,j,\mu,\nu} C_{\mathrm{O}(i,\mu)}\frac{-J_i\uprm{H}}{2}(-D_{(i,\mu),(j,\nu)}(0))\frac{-J_j\uprm{H}}{2}C_{\mathrm{O}(j,\nu)}\nonumber \\
 & -\hbar^2 a\sum_{i,j,\mu,\nu} C_{\mathrm{I}(i,\mu)}\frac{-J_i\uprm{H}}{2}(-D_{(i,\mu),(j,\nu)}(0))\frac{-J_j\uprm{H}}{2}C_{\mathrm{I}(j,\nu)}. \label{eq:decomposedDerivativeSigma}
\end{align}
This quantity plays a role to determine the sign of the inductance as can be seen from Eq.~\eqref{eq:inductance}.
For the case of the canted-AFM state, the Berry phase defined in Eq.~\eqref{eq:emergentE} vanishes because this model has only two sites in a magnetic unit cell, leading to vanishing emergent electric field.
Indeed purple curves in Fig.~\ref{fig:Inductance} show that the inductance becomes negative for the canted-AFM state, indicating that only the interband term contributes to $L$. 
For the case of the other models, the Fermi surface terms can have finite values, and the inductance can be positive unless the chemical potential is in a gap.
As shown in Figs.~\ref{fig:Inductance}(a,b), $\mathrm{Im}[\partial_\omega \Sigma] _{\omega=0}$ becomes negative at the region where $\sigma\smrm{dc}\simeq 0$.

Figure~\ref{fig:Inductance}(c) shows the inductance $L$.
The absolute values of the $L$ show diverging behaviors as the DC conductivities become small because $L\propto \sigma\smrm{dc}^{-2}$.
However, even if the absolute value of $L$ is large in the region $\sigma\smrm{dc}\simeq 0$, this is not good for a practical use because the induction in metals is desirable for application.
On the other hand, the maximum values of the $L$ of the fan state ($\mu \simeq -1.8t_0$) and the 1D kagome model ($\mu \simeq 1.1t_0$) are larger than or comparable to that of the $L$ of the helical state, and both have a finite $\sigma\smrm{dc}$ at the chemical potential where $L$ reaches the maximum value.
This indicates that non-helical magnets, including fan and 1D kagome orders, can potentially provide more efficient inductors than helical magnets.

\begin{figure*}
  \includegraphics[width=\linewidth]{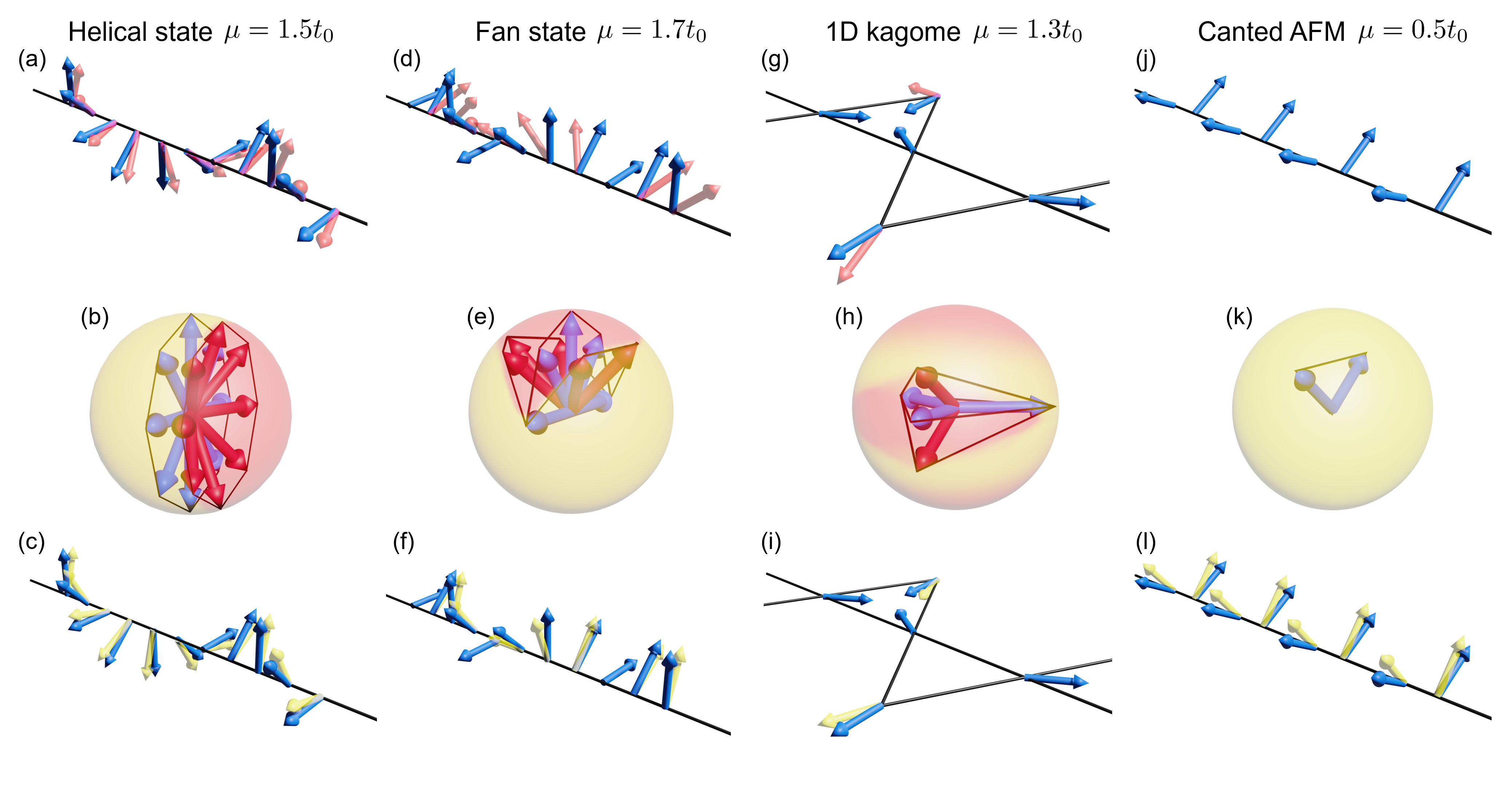}
  \caption{Schematic pictures of the spin dynamics of (a-c) the helical state, (d-f) the fan state, (g-i) the 1D kagome model, and (j-l) the canted-antiferromagnetic state with the Rice-Mele model. Blue spin represents the spin configuration of the ground state of each model, red spin represents the spin configuration fluctuated by $C_{\mathrm{O}(i,x)}$ and $C_{\mathrm{O}(i,y)}$ (a,d,g,j), and yellow spin represents the spin configuration fluctuated by $C_{\mathrm{I}(i,x)}$ and $C_{\mathrm{I}(i,y)}$ (c,f,i,l). The spin configurations of the ground state and the ones fluctuated by $C_{\mathrm{O}(i,x)}$ and $C_{\mathrm{O}(i,y)}$ (a,d,g,j) are projected into unit spheres in (b,e,h,k), respectively. The spins connected by bonds are linked by black lines at their tips, and the solid angle enclosed by the black lines (Berry phase) of the spin configurations fluctuated by $C_{\mathrm{O}(i,x)}$ and $C_{\mathrm{O}(i,y)}$ is approximately indicated in red on the unit sphere.}
  \label{fig:modelsCmove}
\end{figure*}
\subsection{Spin dynamics}\label{subsec:SpinDynamics}

In this section, we present the actual spin dynamics producing the inductance.
Namely, we show the response kernel $C_{\mathrm{O}(i,\mu)}$ and $C_{\mathrm{I}(i,\mu)}$ computed in the itinerant electron system.
Figure~\ref{fig:modelsCmove} shows the schematic pictures of the spin dynamics induced by the AC electric field (or inducing electric current) in the models shown in Fig.~\ref{fig:models} for the chemical potential $\mu=1.5t_0$ (helical state), $\mu=1.7t_0$ (fan state), $\mu=1.3t_0$ (1D kagome model) and $\mu=0.5t_0$ (canted AFM state).
We depict the spin configurations induced by $C_{\mathrm{O}(i,x)}$ and $C_{\mathrm{O}(i,y)}$ in red in Figs.~\ref{fig:modelsCmove}(a,d,g,j) and those induced by $C_{\mathrm{I}(i,x)}$ and $C_{\mathrm{I}(i,y)}$ in yellow in Figs.~\ref{fig:modelsCmove}(c,f,i,l).
We redraw the spin configuration of the ground state of each model in blue for comparison.

The present method captures the real space structure of spin fluctuations. 
To see this, let us compare the spin dynamics of the helical state and that of the other models.
For the helical state that has screw symmetry, we can select a new Bloch basis that consists of two by two Hamiltonians~\cite{Kurebayashi2021}.
In the present method, we redundantly derive the response kernel $C_{\mathrm{O}(i,\mu)}$ and $C_{\mathrm{I}(i,\mu)}$ that have the same value for any site $i$ in the helical state as shown in Figs.~\ref{fig:modelsCmove}(a,c).
This means that, for the helical state, response function $\sigma_{(i,\mu)}(\omega)$ and $\chi_{(i,\mu)}(\omega)$ are spatially uniform in a magnetic unit cell in the rotated frame as derived in Ref.~\cite{Kurebayashi2021}.
In contrast, for the other models that have no screw symmetry, 
the spin fluctuations driven by the AC electric field has a non-uniform component in a magnetic unit cell as shown in Figs.~\ref{fig:modelsCmove}(d,g,j,f,i,l).
In this regard, the present method effectively incorporate non-uniform  components of the response in real space analysis with sublattice structures.
Therefore, the strong point of the present method is its ability to capture the non-uniform components of currents induced by the dynamics of magnetization $\sigma(\omega)$ and the dynamics of the magnetization induced by the electric field $\chi(\omega)$ although the electric field is spatially uniform.
In this way, the present method provides a tool to compute the inductance of the materials that have no screw symmetry.

\begin{figure*}
  \includegraphics[width=\linewidth]{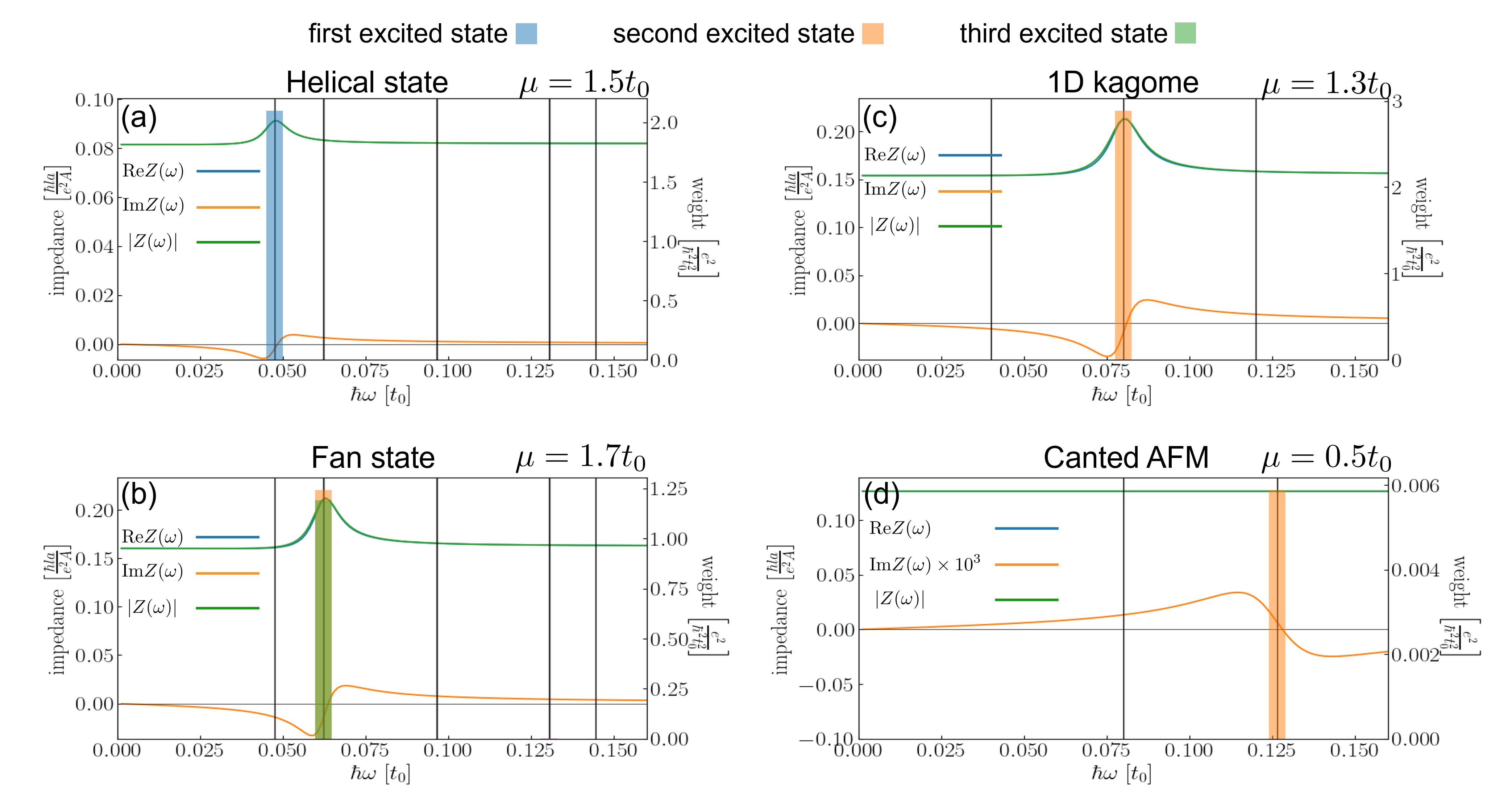}
  \caption{Real part, imaginary part and absolute value of the impedance $Z$ as a function of frequency $\omega$ for (a) the helical state, (b) the fan state, (c) the 1D kagome model, and (d) the canted-antiferromagnetic state with the Rice-Mele model. Black vertical lines indicate eigenenergies of $H\smrm{m}$ and bar graphs on the black lines show the matrix weights of the magnon propagator. We adopt the Gilbert damping: $\alpha=0.1$.}
  \label{fig:impedance}
\end{figure*}

Let us now focus on the response $C_{\mathrm{O}(i,\mu)}$.
Figures~\ref{fig:modelsCmove}(b,e,h,k) show the projection of Figs.~\ref{fig:modelsCmove}(a,d,g,j) into unit spheres.
As mentioned in the Sec.~\ref{subsec:sigmaChi}, the response kernel $C_{\mathrm{O}(i,\mu)}$ is the response of the current induced by the emergent electric field (Eq.~\eqref{eq:emergentE}) in the limit of large Hund coupling,
and the emergent electric field is described by the time variation of the Berry phase in real space induced by the dynamics of the spin configuration.
Red color depicts the approximate Berry phase originating from the spin configuration that is modulated by $C_{\mathrm{O}(i,x)}$ and $C_{\mathrm{O}(i,y)}$.
For the helical state, all the spins tilt perpendicular to the coplane of the spin configuration, and this produces the time variation of the Berry phase as shown in Fig.~\ref{fig:modelsCmove}(b).
On the other hand, for the fan state and the 1D kagome model, the spins move perpendicular to the coplane and in opposite directions when $j\leq N\smrm{site}/2$ and $j>N\smrm{site}/2$ as shown in Figs.~\ref{fig:modelsCmove}(e,h) ($N\smrm{site}=4$ in the 1D kagome model.).
To produce the time variation of the Berry phase in fan state or 1D kagome model, the spins along the same direction must move in opposite directions and perpendicular to the coplane.
Indeed, the spins move in this way by $C_{\mathrm{O}(i,x)}$ and $C_{\mathrm{O}(i,y)}$ as shown in Figs~\ref{fig:modelsCmove}(e,h).
In contrast, in the case of the canted-AFM state that has only two sites in a magnetic unit cell, the Berry phase originating from the spin configuration is always zero, and the emergent electric field does not appear in this system, as shown in Fig.~\ref{fig:modelsCmove}(k).
Note that while the 1D kagome model has the next nearest neighbor hopping, any loop of bonds has no magnetic fluxes, justifying to interpret the Fermi surface term as the emergent electric field.
We also note that, although $J\smrm{H}\sim 0.5t_0$ is not sufficiently large to project the Hilbert space to the spin polarized subspace,
there exists the contribution of the emergent electric field in the spin polarized subspace.

To conclude, this framework captures the spin-motive force even in situations where there is no net induced magnetization by AC field as shown in Figs.~\ref{fig:modelsCmove}(e,h).
This contrasts with the case of helical magnets, where the emergent electric field is regarded as the time variation of the magnetization~\cite{Nagaosa2019}.

\subsection{Impedance}\label{subsec:Impedance}

Figure~\ref{fig:impedance} shows the real part, the imaginary part and the absolute value of the impedance ($\mathrm{Re}Z$, $\mathrm{Im}Z$ and $|Z|$) as a function of frequency $\omega$ at the chemical potentials $\mu$ shown in Fig.~\ref{fig:modelsCmove}.
If the inductance $L$ is positive (negative), $\mathrm{Im}Z$ has a negative (positive) slope at the origin.
This is consistent with the results in Fig.~\ref{fig:Inductance}(c) and the definition of the inductance Eq.~\eqref{eq:inductance}.
Since $C_{\mathrm{O}(i,\mu)}$ and $C_{\mathrm{I}(i,\mu)}$ do not depend on $\omega$, the frequency dependence of impedances are determined by the Green's function $D_{(i,\mu),(j,\nu)}(\omega)$.
We plot the eigenenergies $\epsilon_n$ of $\mathcal{H}\smrm{m}(\bm{k}=0)$ as black vertical lines in Fig.~\ref{fig:impedance}.
If the frequency $\omega$ is sufficiently smaller than the eigenenergies of $H\smrm{m}$, $\mathrm{Im}Z$ are proportional to the frequency as can be seen from Eq.~\eqref{eq:SigmaWithC} with $D(\omega)\sim -H\smrm{m}^{-1}(\bm{k}=0)$.
If the frequency $\omega$ is sufficiently larger than the eigenenergies of $H\smrm{m}$, the impedances converge to constant because $D(\omega)\sim \omega^{-1}(-\sigma_y+i\alpha )^{-1}$.
If the frequency $\omega$ is near the eigenenergies of $H\smrm{m}$, the impedances can have a large $\omega$ dependence.
We now discuss the frequency dependence of the impedances near the eigenenergies of $H\smrm{m}$.
We plot the matrix weights of the magnon propagator $(\hbar\omega -\epsilon_n)^{-1}$ as bar graphs on the black lines in Fig.~\ref{fig:impedance}.
We define matrix weights as
\begin{align}
  W_n= & \Bigg|\sum_{i,j,\mu,\nu} (C_{\mathrm{O}(i,\mu)}+C_{\mathrm{I}(i,\mu)})(v_n)_{(i,\mu)}\nonumber \\
 & \qquad \times (v_n)_{(j,\nu)}^*(-C_{\mathrm{O}(j,\nu)}+C_{\mathrm{I}(j,\nu)})\Bigg|.
\end{align}
Here, $v_n$ is the eigen vector of $\mathcal{H}\smrm{m}(\bm{k}=0)$ associated with the magnon basis.
Namely, we consider the diagonalization of $\mathcal{H}\smrm{m}(\bm{k}=0)$ as
\begin{gather}
  V^\dagger \mathcal{H}\smrm{m}(\bm{k}=0)V=\mathrm{diag}(\{\epsilon_n\}), \\
  -\sigma_y \otimes I_{N\smrm{site}}=(U_{S\to b}\otimes I_{N\smrm{site}})(\sigma_z \otimes I_{N\smrm{site}})(U_{S\to b}^\dagger \otimes I_{N\smrm{site}}),\\
  U_{S\to b}\equiv  \frac{1}{\sqrt{2}}\left(
  \begin{array}{cc}
      1 & 1 \\
      -i & i
    \end{array}
  \right)
\end{gather}
and $v_n$ is the $n$th column vector of $VU_{S\to b}$ where $V$ is the para-unitary matrix satisfying $V^\dagger (-\sigma_y)V=-\sigma_y$.
This $W_n$ is the contribution of the $n$th magnon excitation to the conductivity mediated by magnon.
That is because, in the limit of $\alpha\to 0$,
\begin{align}
 & (-\sigma_y \hbar \omega -\mathcal{H}\smrm{m}(\bm{k}=0))^{-1} \nonumber \\
  = & VU_{S\to b}(\sigma_z \hbar \omega -\sigma_0\otimes \mathrm{diag}(\{\epsilon_n\}))^{-1}U_{S\to b}^\dagger V^\dagger.
\end{align}
If the matrix weight $W_n$ has the finite value, in the limit of $\alpha\to 0$, the conductivity mediated by magnon $\Sigma (\omega)$ diverges and $Z=0$ at $\epsilon_n$.
In practice, because of the finite Gilbert damping $\alpha$, Fig.~\ref{fig:impedance} shows that the impedance shows an enhancement or large modulation with changing $\omega$ only around the eigenenergy $\epsilon_n$ that has finite matrix weight.

Here, we note that the range of the frequency where the impedance shows the large modulation is not necessary to be around the magnon gap.
In the case of the fan state and the 1D kagome model, the spin dynamics induced by AC field  shown in Fig.~\ref{fig:modelsCmove}(b,j,c,k) are orthogonal to the eigenmode of the first magnon excitation modes.
As a result, the AC field does not excite the first magnon excitation modes but the higher magnon excitation modes in the fan state and the 1D kagome model as shown in Fig.~\ref{fig:impedance}(b,c).

\subsection{Q-value}\label{subsec:QValue}
\begin{figure}
  \includegraphics[width=\linewidth]{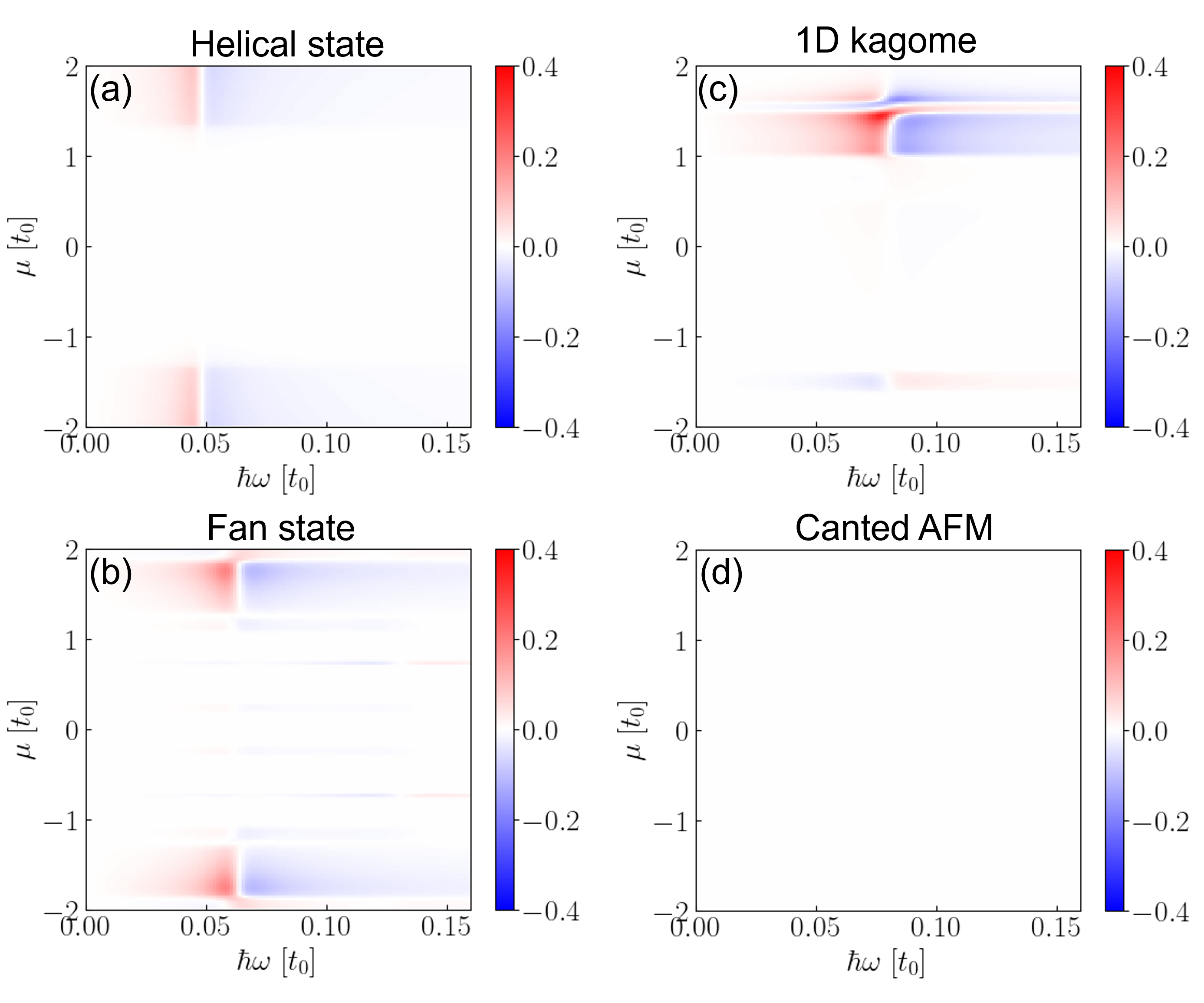}
  \caption{The color plot of Q-values for (a) the helical state, (b) the fan state, (c) the 1D kagome model, and (d) the canted-antiferromagnetic state with the Rice-Mele model.}
  \label{fig:QValues}
\end{figure}
Figure~\ref{fig:QValues} shows the color plot of Q-values as a function of the chemical potential $\mu$ and the frequency $\omega$.
The Q-value is defined as $Q\equiv -\mathrm{Im}Z/\mathrm{Re}Z=\mathrm{Im}\Sigma /(\sigma\smrm{dc}+\mathrm{Re}\Sigma)$.
The absolute value of Q-value becomes large when both the frequency $\omega$ is near the magnon excitation whose matrix weight $W_n$ has a finite value, and the chemical potential is within the region where the absolute value of $\mathrm{Im}[\partial_\omega \Sigma]_{\omega=0}$ (shown in Fig.~\ref{fig:Inductance}(b)) is large.
As for the helical state and the fan state, the Q-value has positive values in the low-frequency region and negative values in the high-frequency regions, respectively.
This mean that $\mathrm{Im}Z$ mostly becomes negative and positive in the low and high-frequency region, which means that the systems are inductive and capacitive, respectively.
These sign changes are due to the sign changes of the imaginary parts of the conductivities mediated by magnon $\mathrm{Im}[\Sigma(\omega)]$ that vary around the magnon excitation.
The 1D kagome model shows the moderate frequency dependence of the Q-value (Fig.~\ref{fig:QValues}(c)) reflecting on the moderate frequency dependence of the impedance (Fig.~\ref{fig:impedance}(c)).
As for the canted-AFM, the Q-value is almost zero because $\mathrm{Re}\Sigma(\omega),\mathrm{Im}\Sigma (\omega)\ll \sigma\smrm{dc}$.
For any material, the Q-value can be large when $\Sigma(\omega)$ has a large imaginary part.
Therefore, the materials whose magnon lifetime is large (i.e. $\alpha \sim 0$) may show large Q-values.

\section{Discussion}\label{sec:Discussion}
\subsection{Candidate materials and application}\label{subsec:CandidateMaterials}
The present method can be applied to various magnetic materials.
We now discuss candidate materials, starting with those described by the models that show the large inductance in metallic state as shown in Fig.~\ref{fig:Inductance}(c) i.e. the fan state and the 1D kagome model.
For fan states, $\mathrm{Gd}_3\mathrm{Ru}_4\mathrm{Al}_{12}$~\cite{Chandragiri2016,Nakamura2018,Hirschberger2019} and $\mathrm{Y}_{1-x}\mathrm{Tb}_x\mathrm{Mn}_6\mathrm{Sn}_6$~\cite{Venturini1996,Wang2021} are candidate materials because they show the fan phase under magnetic fields.
For 1D kagome models, $\mathrm{Mn}_3\mathrm{Sn}$~\cite{Tomiyoshi1982} and $\mathrm{Mn}_3\mathrm{Ir}$~\cite{Izumi1999} are candidate materials because they show the antiferromagnetic triangular spin structure at room temperature. 
Figure~\ref{fig:impedance}(a) indicates the imaginary part of complex resistance $|\mathrm{Im} \rho| \SI{\sim 1}{\micro\ohm\centi\meter}$ for the helical state with a lattice constant of $a\sim \SI{0.5}{\nano\meter}$ and this value is consistent with the reported values in Ref.~\cite{Kitaori2021,Kitaori2023,Kitaori2024}.
Assuming the same parameters for the fan state and the 1D kagome model, the imaginary part of complex resistance shown in Figs.~\ref{fig:impedance}(b,c) is estimated to be $|\mathrm{Im} \rho| \sim \SI{5}{\micro\ohm\centi\meter}$ and $|\mathrm{Im} \rho| \sim \SI{15}{\micro\ohm\centi\meter}$, respectively, which are expected to be measurable in the above materials.

When selecting the suitable materials, there are several points to be noted.
First, the formulation in Secs.~\ref{sec:formalism} and \ref{sec:InductanceAndImpedance} is invariant under global spin transformations because of the absence of SOC or anisotropy in spin space.
Accordingly, the models that exhibit the same spin structure by global spin transformations will give the same inductance.
Therefore, $\mathrm{Mn}_3\mathrm{Sn}$ and $\mathrm{Mn}_3\mathrm{Ir}$ are good candidate materials for the 1D kagome model although the spin structures of these materials are not exactly the same as the spin structure shown in Fig.~\ref{fig:models}(c).
Second, the precise expression of the spin Hamiltonian is not important for the prediction of the inductance.
The important point is that the response kernel of the magnetization $D(\omega)$ in Eq.~\eqref{eq:Sigma} has a diverging behavior around the magnetic excitation.
Therefore, the spin Hamiltonian of the materials do not have to be the same as that presented in Sec.~\ref{subsec:1Dmodels} and, moreover, the present method can be applied if the classical spin configuration is stable and consequently the Lagrangian is described by the bilinear form of the localized-spin perturbations $\tilde{S}_j^{\mu=x,y}$.
Thus, $\mathrm{Mn}_3\mathrm{Sn}$ and $\mathrm{Mn}_3\mathrm{Ir}$ are good candidate materials for the 1D kagome model although the antiferromagnetic triangular spin structure is not realized as a ground state. 
Also, $\mathrm{Gd}_3\mathrm{Ru}_4\mathrm{Al}_{12}$ and $\mathrm{Y}_{1-x}\mathrm{Tb}_x\mathrm{Mn}_6\mathrm{Sn}_6$ are good candidate materials although the fan phase in these materials is not described by the ground state of Eq.~\eqref{eq:fanHm}.

There are many spin structures that may exhibit large inductance other than the spin structures shown in Fig.~\ref{fig:models}, and the present method can be applied to magnetic materials where classical spin configurations are realized and the localized spins have sufficiently large magnetic moments.
Additionally, a noncollinear spin structure with more than two sites in a magnetic unit cell is essential for generating a finite emergent electric field.
One of the candidate materials satisfying these conditions is $\mathrm{Mn}_5\mathrm{Sn}_3$, which shows non-collinear spin structure below \SI{62}{\K}~\cite{Brown1992,Gottschilch2012}.
Another candidate material is $\mathrm{EuIn}_2\mathrm{As}_2$~\cite{Soh2023,Riberolles2021,Regmi2020,Sato2020} that shows the broken-helix spin structure below \SI{16}{\K}.
Furthermore, from an industrial application perspective, $\mathrm{Mn}_5\mathrm{Sn}_3$ and $\mathrm{Mn}_3\mathrm{Sn}$ may be suitable materials for the inductor application in that the elements of these materials are easy to obtain compared to the elements of other materials mentioned above.

When computing the inductance of the actual materials by using present method, there are several points to consider.
The present method focuses on the response mediated by the dynamics of the magnetization which is described by the Feynman diagram shown in Fig.~\ref{fig:FeynmanDiagram}.
There are actually other contributions to the response function, such as the self-energy (i.e. Hartree terms and Fock terms by the magnon scattering) and vertex correction by the magnon scattering, which are the same order of the perturbation as the Feynman diagram shown in Fig.~\ref{fig:FeynmanDiagram}.
Although the contribution from the diagram shown in Fig.~\ref{fig:FeynmanDiagram} is dominant due to the diverging behaviors of the $\bm{k}=0$ component of the magnon Green's function around the magnon excitation, for applying the present method to actual materials, the calculation including the self-energy and vertex correction gives more precise prediction.

\subsection{Comparison with other methods and outlook}\label{subsec:Comparison}
In this section, we compare the present method with other methods and discuss the future outlook.
The present method captures the inductance originating from dynamics of non-collinear spin structures.
In contrast, there are methods that capture the inductance originating from the SOC or the U(1) gauge fluctuation.

It is theoretically shown that SOC induces ``spin-orbit emergent inductance'' even in ferromagnets~\cite{Ieda2021,Yamane2022,Araki2023}.
Specifically, the inductance mediated by the dynamics of a ferromagnet with SOC is derived in continuum models: parabolic dispersion with SOC as in Refs.~\cite{Ieda2021,Yamane2022} and the Dirac electron that originates from SOC as in Ref~\cite{Araki2023}.
Reference~\cite{Araki2023} treats a heterostructure of ferromagnet and Dirac system that is made of the surface state of a topological insulator and also discusses the application to magnetic structures other than ferromagnets.
In the present formalism, SOC can be introduced by adding $-\sum_{i,j,\tau}c_{(i,j)}^\dagger i\bm{\lambda}\cdot \bm{\sigma} c_{(i,j)+\tau}+h.c.$ to Eq.~\eqref{eq:HelLab}, resulting in
\begin{align}
  &t_{j,\tau}\tilde{c}_{(i,j)}^\dagger U_j^\dagger U_{j+\tau} \tilde{c}_{(i,j)+\tau}\nonumber \\
  \to &\sqrt{t_{j,\tau}^2+|\bm{\lambda}|^2}\tilde{c}_{(i,j)}^\dagger U_j^\dagger e^{i\theta \frac{\bm{\lambda}}{|\bm{\lambda}|}\cdot \bm{\sigma}} U_{j+\tau} \tilde{c}_{(i,j)+\tau} \label{eq:SOCchange}
\end{align}
within Eq.~\eqref{eq:HelFerro}, where $\theta=\arctan(|\bm{\lambda}|/t_{j,\tau})$.
This means that the effective Peierls phase is introduced by the SOC even in ferromagnets, where $U_j=\sigma_0$ for any site $j$.
In the presence of SOC, the emergent electric field Eq.~\eqref{eq:emergentE} is rewritten as
\begin{align}
  \frac{\hbar }{eN\smrm{site}a}\partial_t \mathrm{Im}\log \prod_j \mel{\bm{n}_j}{e^{i\theta \frac{\bm{\lambda}}{|\bm{\lambda}|}\cdot \bm{\sigma}}}{\bm{n}_{j+1}}.
\end{align}
Therefore, the SOC can be interpreted as the effective spin structure producing the emergent electric field~\cite{Yamane2022}.
We show the results of numerical calculations of the inductance in the ferromagnetic state and antiferromagnetic state with SOC in Appendix~\ref{app:spinorbit}. 
In the ferromagnetic state, results similar to those in the helical state are obtained.

There is also a theory for the inductance in the strongly-correlated system~\cite{Oh2024}.
Starting from the t--J model, they consider the current coupling to U(1) gauge field of holon and spinon and obtain the inductance from the conductivity.
This method can be applicable to the strongly-correlated magnets which are not described by the classical spin configuration,
whereas the applicability of the present method is limited to the classical spin configuration.

The present method can also be applied to derive Hall inductance in 2D and 3D systems.
For 1D systems, the inductance can be decomposed into positive component from the Fermi surface terms and negative component from the interband effect terms.
The former component does not depend on $\tau$ and the latter component is proportional to the inverse square of the relaxation time $\tau^{-2}$ as shown in Eq.~\eqref{eq:decomposedL} (note that $\sigma\smrm{dc}\propto \tau$ and $C_{\mathrm{O}(i,\mu)}^\lambda \propto \tau$).
In contrast, in the Hall inductance of 2D or 3D systems, the cross terms proportional to the inverse of $\tau$ (e.g. $C_{\mathrm{O}(i,\mu)}^xC_{\mathrm{I}(j,\nu)}^y$) appears,
although the decomposition is no longer valid.
This may be detectable by varying the impurity doping.
Another interesting point for the 2D or 3D systems is the effect of time-varying emergent magnetic flux.
The emergent electric field is described by the time variation of the Berry phase induced by the dynamics of the spin configuration as shown in Eq.~\eqref{eq:emergentE} and Fig.~\ref{fig:modelsCmove}(b,e,h,k) and this Berry phase is interpreted as the magnetic flux surrounded by a coil.
In addition to the Berry phase, the emergent magnetic flux appears in 2D or 3D systems even in the condition that hoppings have finite values only between nearest neighbor sites.
Therefore, although we can not perform the twisted parallel-transport gauge transformation Eq.~\eqref{eq:twistedGaugeTrans} for the system with the emergent magnetic flux,
the time variation of the emergent magnetic flux may induce the emergent electric field and the emergent inductance.
In the context of DC currents response under the static electric field, the effect of the emergent magnetic flux is studied as the topological Hall effect~\cite{Bruno2004,Nagaosa2013} and the skyrmion Hall effect~\cite{Zang2011}.
Therefore, one may extend the analysis of the emergent electric field in Sec.~\ref{subsubsec:CO} to 2D and 3D systems incorporating the effects of the emergent magnetic flux, which can lead to an intuitive understanding of the dynamics of the emergent magnetic flux on the emergent inductance and the motion of skyrmions under the AC field applied in different directions with phase differences.

\acknowledgements
We thank Naoto Nagaosa, Joji Nasu, Shun Okumura, and Aki Kitaori for fruitful discussions.
This work was supported by JST SPRING, Grant Number JPMJSP2108 (T.A.) and JSPS KAKENHI Grant 23K25816, 23K17665, 24H02231 (T.M.).

\appendix

\section{Magnon operator formalism}\label{app:magnonFormalism}
In this section, we derive the emergent inductance using magnon operator instead of performing the coordinate transformation of the itinerant electron system (Eq.~\eqref{eq:U}) and show the equivalence of the results, especially, the fact that longitudinal inductance has positive component due to the Fermi surface terms and negative component due to the polarization.
We start from the Hamiltonian in the real space, Eqs.~\eqref{eq:HelLab}, \eqref{eq:HcLab} and \eqref{eq:HmLab}.
First, we perform the Holstein-Primakoff transformation,
\begin{align}
 & T_j \tilde{S}_{(i,j)} =S_{(i,j)}, \\
 & T_j = \left(
  \begin{array}{ccc}
      \cos \phi_j \cos \theta_j & -\sin \phi_j & \cos \phi_j \sin \theta_j \\
      \sin \phi_j \cos \theta_j & \cos \phi_j & \sin \phi_j \sin \theta_j \\
      -\sin \theta_j & 0 & \cos \theta_j \\
    \end{array}
  \right), \\
 & \left(
  \begin{array}{c}
      \tilde{S}_{(i,j)}^x \\ \tilde{S}_{(i,j)}^y\\ \tilde{S}_{(i,j)}^z
    \end{array}
  \right)=\hbar \left(
  \begin{array}{c}
      \frac{\sqrt{2S}}{2}(b_{(i,j)}+b_{(i,j)}^\dagger) \\
      \frac{\sqrt{2S}}{2i}(b_{(i,j)}-b_{(i,j)}^\dagger) \\
      S-b_{(i,j)}^\dagger b_{(i,j)}
    \end{array}
  \right),
\end{align}
and obtain the Hamiltonian using the magnon operators $b_{(i,j)}$ as follows:
\begin{align}
  H\smrm{el} = & -\sum_{i,j,\tau} t_{j,\tau}c_{(i,j)}^\dagger \sigma_0 c_{(i,j)+\tau}+h.c.\nonumber \\
 & \qquad   -\sum_{i,j} \frac{J_j\uprm{H}\hbar S}{2}\bm{n}_j \cdot      c_{(i,j)}^\dagger \bm{\sigma} c_{(i,j)} \\
  H\smrm{c}  = & -\sum_{i,j} \frac{J_j\uprm{H}\hbar \sqrt{2S}}{4}b_{(i,j)}\cdot c_{(i,j)}^\dagger U_j \sigma_- U_j^\dagger c_{(i,j)}+h.c., \\
  H\smrm{m}  = & -\sum_{\substack{i,j,\tau, \\\mu=x,y,z,\\\nu=x,y,z}} \tilde{J}_{j,\tau}^{\mu\nu}\tilde{S}_{(i,j)}^\mu \tilde{S}_{(i,j)+\tau}^\nu-\sum_{i,j}\tilde{\bm{B}}_j\cdot \tilde{\bm{S}}_{(i,j)} \nonumber \\
  = & -\sum_{\substack{i,j,\tau, \\\mu=x,y,\\\nu=x,y,\\\mu'=\pm,\\ \nu'=\pm}} \hbar ^2 S\tilde{J}_{j,\tau}^{\mu\nu}[U_{S\to b}^\dagger]_{\mu'}^\mu (\Psi_{(i,j)}^{\mu'})^\dagger [U_{S\to b}]_{\nu'}^\nu \Psi_{(i,j)+\tau}^{\nu'}\nonumber\\
 & +\sum_{\substack{i,j,\tau \\ \mu'=\pm}} \hbar ^2 S\frac{1}{2}(\tilde{J}_{j,\tau}^{zz} +\tilde{J}_{j-\tau,\tau}^{zz} )(\Psi_{(i,j)}^{\mu'})^\dagger \Psi_{(i,j)}^{\mu'}\nonumber \\
 & +\sum_{i,j, \mu'=\pm} \hbar ^2 S\frac{1}{2}\frac{\tilde{B}_j^z}{\hbar S}(\Psi_{(i,j)}^{\mu'})^\dagger \Psi_{(i,j)}^{\mu'}+const., \\
  \Psi_{(i,j)}^{\mu'}\equiv & \begin{cases}
                                b_{(i,j)} & \cdots \mu'=+ \\
                                b_{(i,j)}^\dagger & \cdots \mu'=- \\
                              \end{cases}, \\
  U_{S\to b}\equiv & \frac{1}{\sqrt{2}}\left(
  \begin{array}{cc}
      1 & 1 \\
      -i & i
    \end{array}
  \right),
\end{align}
where $\sigma_\pm=(\sigma_x\pm i\sigma_y)$, and $\tilde{J}_{j,\tau}=T_j J_{j,\tau}T_{j+\tau}\uprm{T}$.
Here, we ignore $O(b^2c^2)$ terms in $H\smrm{Hund}$ and $O(b^4)$ terms in $H\smrm{m}$.
Note that we also drop the terms proportional to $\tilde{S}^z\tilde{S}^{\mu=x,y}$ and $\tilde{S}^{\mu=x,y}$ which produce only single magnon terms or $O(b^3)$ terms because $\{\bm{n}_j\}$ is the spin configuration of the ground state of $H\smrm{m}$.

Next, we perform the Fourier transformation of the fermion operators and magnon operators and obtain the Hamiltonian in the momentum space as follows:
\begin{align}
  H\smrm{el} = & -\sum_{\bm{k},j,\tau} t_{j,\tau}c_{\bm{k},j}^\dagger \sigma_0 c_{\bm{k},j+\tau}e^{i\bm{k}\cdot \bm{r}_\tau}+h.c.\nonumber \\
 & \qquad   -\sum_{\bm{k},j} \frac{J_j\uprm{H}\hbar S}{2}\bm{n}_j \cdot      c_{\bm{k},j}^\dagger \bm{\sigma} c_{\bm{k},j} \\
  H\smrm{c}  = & \frac{1}{\sqrt{N}}\sum_{\substack{\bm{k},\bm{q},j, \\ \mu'=\pm}} \frac{-J_j\uprm{H}}{2}\Psi_{\bm{q},j}^{\mu'}\cdot c_{\bm{k}+\bm{q},j}^\dagger \mathcal{M}^{-\mu'} c_{\bm{k},j}, \\
  \mathcal{M}^{\mu'}\equiv & \hbar \frac{\sqrt{2S}}{2}U_j \sigma_{\mu'} U_j^\dagger, \\
  H\smrm{m}= & \sum_{\substack{\bm{k},j,\tau, \\\mu',\nu'}}\mathcal{H}_{(j,\mu'),(j+\tau,\nu')}\uprm{mag}(\bm{k})(\Psi_{\bm{k},j}^{\mu'})^\dagger \Psi_{\bm{k},j+\tau}^{\nu'}\nonumber \\
  &+const.,\label{eqApp:HmBilinear}
\end{align}
with
\begin{align}
 \mathcal{H}_{(j,\mu'),(j+\tau,\nu')}\uprm{mag}(\bm{k})
  &\equiv  \sum_{\substack{\mu=x,y, \\\nu=x,y}} \hbar ^2 Se^{i\bm{k}\cdot \bm{r}_\tau}[U_{S\to b}^\dagger]_{\mu}^{\mu'}\nonumber \\
 & \qquad\times \Big[-\tilde{J}_{j,\tau}^{\mu\nu} \nonumber \\
 &\qquad \quad +\frac{1}{2}\delta_{\mu,\nu}\delta_{\tau,0}\sum_{\tau'}(\tilde{J}_{j,\tau'}^{zz} +\tilde{J}_{j-\tau',\tau'}^{zz} )\nonumber \\
 &\qquad \quad +\frac{1}{2}\delta_{\mu,\nu}\delta_{\tau,0}\frac{\tilde{B}_j^z}{\hbar S}\Big]\nonumber \\
 &\qquad \times [U_{S\to b}]_{\nu'}^\nu \nonumber \\
 &=  \sum_{\substack{\mu=x,y, \\\nu=x,y}} \hbar ^2 S[U_{S\to b}^\dagger]_{\mu}^{\mu'}[\mathcal{H}\smrm{m}(\bm{k})]_\nu^\mu[U_{S\to b}]_{\nu'}^\nu.\label{eqApp:HmMag}
\end{align}
As can be seen from Eq.~\eqref{eqApp:HmMag}, $\mathcal{H}_{(j,\mu'),(j+\tau,\nu')}\uprm{mag}(0)$ is transformed from the coefficient matrix of $H\smrm{m}$ in Eq.~\eqref{eq:HmBilinear} by the unitary transformation $U_{S\to b}$.
Note that we can assume $\mathrm{Im}\tilde{J}_{j,0}^{xy}=\mathrm{Im}\tilde{J}_{j,0}^{yx}=0$ because
\begin{align}
  i[\tilde{S}_j^x,\tilde{S}_j^y]\simeq & \frac{\hbar^2 S}{2}[b_j+b_j^\dagger,b_j-b_j^\dagger]\nonumber \\
  = & -\hbar^2 S[b_j,b_j^\dagger]=-\hbar ^2 S \label{eq:commutatorOfSxSy} \\
  \tilde{J}_{j,0}^{xy}\tilde{S}_j^{x}\tilde{S}_j^{y}+ \tilde{J}_{j,0}^{yx}\tilde{S}_j^{y}\tilde{S}_j^{x}
  = & \mathrm{Re}\tilde{J}_{j,0}^{xy}\{S_j^x,S_j^y\}+i\mathrm{Im}\tilde{J}_{j,0}^{xy}[S_j^x,S_j^y]\nonumber \\
  = &\mathrm{Re}\tilde{J}_{j,0}^{xy}\{S_j^x,S_j^y\}+ const. \label{eq:J0xy}
\end{align}
Here, we use the Hermicity of $H\smrm{m}$ in the first line of Eq.~\eqref{eq:J0xy}.
By assuming $\mathrm{Im}\tilde{J}_{j,0}^{xy}=\mathrm{Im}\tilde{J}_{j,0}^{yx}=0$, we obtain $\tilde{J}_{j,\tau}^{\mu\nu},\tilde{B}_j^z\in \mathbb{R}$ because of the Hermicity of $H\smrm{m}$ and this assumption establishes the particle-hole symmetry of $\mathcal{H}\uprm{mag}$ Eq.~\eqref{eqApp:symmetryOfHmag} and $\mathcal{H}\smrm{m}^\mathrm{T}(0)=\mathcal{H}\smrm{m}(0)$ used in Eq.~\eqref{eq:decomposedL}.
Therefore, we ignore $\mathrm{Im}\tilde{J}_{j,0}^{xy}$ and $\mathrm{Im}\tilde{J}_{j,0}^{yx}$ hereafter.

Second, we consider the linear response of the current density induced by the AC electric field $j\smrm{m}$.
By using the Green's function formalism, we consider the following response function shown in Fig.~\ref{fig:FeynmanDiagram},
\begin{align}
  \Sigma^{\lambda\rho}(i\omega_m) & =\sum_{i,j,\mu',\nu'}\sigma_{(i,-\mu')}^\lambda(i\omega_m) D_{(i,\mu'),(j,\nu')}\uprm{mag}(i\omega_m)\nonumber \\
 & \qquad \times \frac{-J_j\uprm{H}}{2} \chi_{(j,\nu')}^\rho (i\omega_m) \frac{1}{-\omega_m}\label{eqApp:Sigma} \\
  \sigma_{(i,\mu')}^\lambda(i\omega_m) & \equiv-\frac{1}{\beta V}\sum_{\bm{k},\omega_l}\mathrm{Tr}[j_{\bm{k}}^\lambda \mathcal{G}_{\bm{k}}(i\omega_l+i\omega_m)\nonumber \\
 & \qquad \qquad\qquad\times (\Pi_i\mathcal{M}^{-\mu'})\mathcal{G}_{\bm{k}}(i\omega_l)]\frac{-J_i\uprm{H}}{2} \label{eqApp:sigmaIMu} \\
  \chi_{(j,\nu')}^\rho (i\omega_m) & \equiv -\frac{1}{\beta N}\sum_{\bm{k},\omega_n}\mathrm{Tr}[ (\Pi_j\mathcal{M}^{-\nu'})\mathcal{G}_{\bm{k}}(i\omega_n +i\omega_m)\nonumber \\
 & \qquad \qquad\qquad \times (-j_{\bm{k}}^\rho)\mathcal{G}_{\bm{k}}(i\omega_n)] \label{eqApp:chiJMu}
\end{align}
with
\begin{align}
  D(\omega)\uprm{mag}=(\sigma_z \hbar \omega +i\alpha\hbar \omega -\mathcal{H}\uprm{mag}(\bm{k}=0))^{-1},
\end{align}
where $\sigma_z$ is the Pauli matrix acting on the two dimensional space spanned by $\mu'=\pm$.
Note the sign of the the index $\mu'$ in $\sigma_{(i,-\mu')}^\lambda$ in Eq.~\eqref{eqApp:Sigma} due to the Hermitian conjugate operator $(\Psi_{\bm{k}=0,j}^{\mu'})^\dagger$ in $H\smrm{m}$ in Eq.~\eqref{eqApp:HmBilinear}.

Third, we derive the $\sigma_{(i,\mu'=\pm)}^\lambda$ and $\chi_{(j,\nu'=\pm)}^\rho$.
Similarly to Eqs.~\eqref{eq:sigmaIMuWithC} and \eqref{eq:chiJNuWithC}, $\sigma_{(i,\mu')}^\lambda$ and $\chi_{(j,\nu')}^\rho$ are given as
\begin{widetext}
  \begin{gather}
    \sigma_{(i,\mu')}^\lambda       (\omega_m)    =i\hbar \omega \Gamma_{\mathrm{O}(i,\mu')}^\lambda\frac{-J_i\uprm{H}}{2}+i\hbar \omega \Gamma_{\mathrm{I}(i,\mu')}^\lambda\frac{-J_i\uprm{H}}{2}\label{eqApp:sigmaIMuWithC} \\
    \chi_{(j,\nu')}^\rho(\omega)=-i\hbar\omega v \Gamma_{\mathrm{O}(j,\nu')}^\rho+i\hbar \omega v \Gamma_{\mathrm{I}(j,\nu')}^\rho, \label{eqApp:chiJNuWithC}\\
    \Gamma_{\mathrm{O}(i,\mu')}^\lambda  =\frac{\tau}{\hbar }\sum_{u}\int_{\mathrm{BZ}} \frac{d\bm{k}}{(2\pi)^d} \mel{u}{{j}_{\bm{k}}^\lambda}{u}\mel{u}{\Pi_i\mathcal{M}^{-\mu'}}{u}f_0'(\xi_{\bm{k},u}) \label{eqApp:CO} \\
    \Gamma_{\mathrm{I}(i,\mu')}^\lambda  =-i\sum_{u\neq v}\int_{\mathrm{BZ}} \frac{d\bm{k}}{(2\pi)^d}\frac{\mel{u}{{j}_{\bm{k}}^\lambda}{v}\mel{v}{\Pi_i\mathcal{M}^{-\mu'}}{u}-\mel{u}{\Pi_i\mathcal{M}^{-\mu'}}{v}\mel{v}{{j}_{\bm{k}}^\lambda}{u}}{(\xi_{\bm{k},u}-\xi_{\bm{k},v})^2}f_0(\xi_{\bm{k},u}).  \label{eqApp:CI}
  \end{gather}
\end{widetext}
Note that $\Gamma_{\mathrm{O}(i,\mu')}^\lambda$ and $\Gamma_{\mathrm{I}(i,\mu')}^\lambda$ are not real numbers in general because $\mathcal{M}^{\mu'}$ is not Hermitian in contrast to $C_{\mathrm{O}(i,\mu)},\ C_{\mathrm{I}(i,\mu)} \in \mathbb{R}$ in Eqs.~\eqref{eq:CO} and \eqref{eq:CI}.
Instead, there are the relations between $\Gamma_{(i,\mu')}^\lambda$ and $\Gamma_{(i,-\mu')}^\lambda$ as follows:
\begin{align}
  \Gamma_{\mathrm{O}(i,\mu')}^\lambda & =\Gamma_{\mathrm{O}(i,-\mu')}^{\lambda*}, \label{eqApp:relationGammaO} \\
  \Gamma_{\mathrm{I}(i,\mu')}^\lambda & =\Gamma_{\mathrm{I}(i,-\mu')}^{\lambda*}.\label{eqApp:relationGammaI}
\end{align}

Finally, we evaluate the conductivity mediated by magnon $\Sigma^{\lambda\rho}(i\omega_m)$, and the inductance $L^{\lambda\rho}$.
By using Eqs.~\eqref{eqApp:Sigma}, \eqref{eqApp:sigmaIMuWithC} and \eqref{eqApp:chiJNuWithC}, we obtain the conductivity mediated by magnon as
\begin{align}
  \Sigma^{\lambda \rho} (\omega)= & i\hbar ^2 \omega v \sum_{i,j,\mu',\nu'} (\Gamma_{\mathrm{O}(i,-\mu')}^\lambda+\Gamma_{\mathrm{I}(i,-\mu')}^\lambda)\frac{-J_i\uprm{H}}{2}\nonumber \\
 & \qquad \times D_{(i,\mu'),(j,\nu')}\uprm{mag}(\omega)\frac{-J_j\uprm{H}}{2}(-\Gamma_{\mathrm{O}(j,\nu')}^\rho+\Gamma_{\mathrm{I}(j,\nu')}^\rho).
\end{align}
By the definitions of the impedance $Z^{\lambda \rho} (\omega)\equiv (l/A)(\sigma \smrm{dc}^{\lambda \rho}+\Sigma^{\lambda \rho}  (\omega))^{-1}$ and the inductance $L^{\lambda \rho} \equiv -\mathrm{Im}[\partial_\omega Z (\omega)]_{\omega=0}$, we obtain
\begin{align}
  L^{\lambda \rho} = & \frac{\hbar ^2l v }{A(\sigma\smrm{dc}^{\lambda \rho})^2}  \sum_{i,j,\mu',\nu'} (\Gamma_{\mathrm{O}(i,-\mu')}^\lambda+\Gamma_{\mathrm{I}(i,-\mu')}^\lambda)\frac{-J_i\uprm{H}}{2}\nonumber \\
 & \qquad \times D_{(i,\mu'),(j,\nu')}\uprm{mag}(0)\frac{-J_j\uprm{H}}{2}(-\Gamma_{\mathrm{O}(j,\nu')}^\rho+\Gamma_{\mathrm{I}(j,\nu')}^\rho).\label{eqApp:L}
\end{align}

Now, we focus on the longitudinal inductance $L^{\lambda\lambda}$.
Using the fact that $\tilde{J}_{j,\tau}^{\mu\nu},\tilde{B}_j^z\in \mathbb{R}$ and Eq.~\eqref{eqApp:HmMag}, we obtain the particle-hole symmetry of $\mathcal{H}\uprm{mag}(0)$ and $D\uprm{mag}(0)$ as
\begin{align}
  \sigma_x \mathcal{H}^{\mathrm{mag}*}(0)\sigma_x=\mathcal{H}\uprm{mag}(0)\label{eqApp:symmetryOfHmag} \\
  \sigma_x D^\mathrm{magT}(0) \sigma_x =D\uprm{mag}(0).  \label{eqApp:symmetryOfD}
\end{align}
Because of Eqs.~\eqref{eqApp:symmetryOfD}, \eqref{eqApp:relationGammaO} and \eqref{eqApp:relationGammaI}, the cross terms cancel out in Eq.~\eqref{eqApp:L} if $\lambda=\rho$.
Therefore, the inductance $L^{\lambda\lambda}$ is decomposed as
\begin{widetext}
  \begin{align}
    L^{\lambda \lambda} = & \frac{\hbar ^2l v }{A(\sigma\smrm{dc}^{\lambda \lambda})^2}  \sum_{i,j,\mu',\nu'} \Gamma_{\mathrm{O}(i,-\mu')}^\lambda\frac{-J_i\uprm{H}}{2}(-D_{(i,\mu'),(j,\nu')}\uprm{mag}(0))\frac{-J_j\uprm{H}}{2}\Gamma_{\mathrm{O}(j,\nu')}^\lambda \nonumber \\
 & -\frac{\hbar ^2l v }{A(\sigma\smrm{dc}^{\lambda \lambda})^2}  \sum_{i,j,\mu',\nu'} \Gamma_{\mathrm{I}(i,-\mu')}^\lambda\frac{-J_i\uprm{H}}{2}(-D_{(i,\mu'),(j,\nu')}\uprm{mag}(0))\frac{-J_j\uprm{H}}{2}\Gamma_{\mathrm{I}(j,\nu')}^\lambda. \label{eqApp:Ldecomposed}
  \end{align}
\end{widetext}
Here, the eigenenergies of $\mathcal{H}\uprm{mag}(0)$ must be positive because we assume that $\{\bm{n}_j\}$ is the spin configuration of the ground state of $H\smrm{m}$ and there is a finite gap to the first excited state.
Therefore, $\mathcal{H}\uprm{mag}(0)$ is positive definite because of Sylvester's law of inertia and thus $-D\uprm{mag}(0)$ is positive definite.
Hence, the first term and the second term in Eq.~\eqref{eqApp:Ldecomposed} are positive and negative, respectively.
In this way, we can derive the formalism without the coordinate transformation of the itinerant electron system that gives the same result to the method described in Sec.~\ref{sec:formalism}.
We note that 
negative contribution to $L^{\lambda\lambda}$ may seem to be energetically unstable as the inductance $L$ has the energy $LI^2/2$ under the current $I$, but 
such contribution can be obtained from the equivalent circuit model with a capacitor as shown in Fig.~\ref{fig:classicalCircuit}.
Similar argument on energetic instability is performed in terms of phenomenological classical circuit model~\cite{Furuta2023,Furuta2024}.

\section{Derivation of $C_{\mathrm{O}(i,\mu)}^\lambda$ and $C_{\mathrm{I}(i,\mu)}^\lambda$}\label{app:Cderivation}
In this section, we derive Eq.~\eqref{eq:sigmaIMuWithC}.
We start from Eq.~\eqref{eq:sigmaIMu} and rewrite it in continuum $\bm{k}$-space as
\begin{widetext}
\begin{align}
  \sigma_{(i,\mu)}^\lambda(i\omega_m)=\frac{1}{\beta}\sum_{\omega_l}\int\smrm{BZ}\frac{d\bm{k}}{(2\pi)^d}\mathrm{Tr}[j_{\bm{k}}^\lambda \mathcal{G}_{\bm{k}}(i\omega_l+i\omega_m) (\Pi_i\sigma_\mu)\mathcal{G}_{\bm{k}}(i\omega_l)]\frac{-J_i\uprm{H}}{2} 
\end{align}
\end{widetext}
In the following, we show $K_{(i,\mu)}^\lambda(\omega)=i\hbar \omega C_{\mathrm{O}(i,\mu)}^\lambda+i\hbar \omega C_{\mathrm{I}(i,\mu)}^\lambda$ where $K_{(i,\mu)}^\lambda(\omega)\equiv \sigma_{(i,\mu)}^\lambda(\omega)/\frac{-J_i\uprm{H}}{2} $.

First, performing the analytic continuation $i\omega_l\to \xi$ and $i\omega_m \to \hbar\omega$, we obtain
\begin{align}
  K_{(i,\mu)}^\lambda(\omega)= &-\int_{-\infty}^{\infty} \frac{d\xi}{2\pi i} \int\smrm{BZ}\frac{d\bm{k}}{(2\pi)^d}f_0(\xi)\nonumber \\
  &\qquad \quad \ \times \Big[\mathrm{Tr}[j_{\bm{k}}^\lambda G_{\bm{k}}^R(\xi+\hbar\omega) (\Pi_i\sigma_\mu)G_{\bm{k}}^R(\xi)]\nonumber \\
  & \qquad \qquad- \mathrm{Tr}[j_{\bm{k}}^\lambda G_{\bm{k}}^R(\xi+\hbar\omega) (\Pi_i\sigma_\mu)G_{\bm{k}}^A(\xi)]\nonumber \\
  & \qquad \qquad+ \mathrm{Tr}[j_{\bm{k}}^\lambda G_{\bm{k}}^R(\xi) (\Pi_i\sigma_\mu)G_{\bm{k}}^A(\xi-\hbar\omega)]\nonumber \\
  & \qquad \qquad- \mathrm{Tr}[j_{\bm{k}}^\lambda G_{\bm{k}}^A(\xi) (\Pi_i\sigma_\mu)G_{\bm{k}}^A(\xi-\hbar\omega)]\Big],
\end{align}
where $G_{\bm{k}}^R(\xi)\equiv (\xi-\mathcal{H}\smrm{el}(\bm{k})+ i\Delta)^{-1}$ and $G_{\bm{k}}^A(\xi)\equiv (\xi-\mathcal{H}\smrm{el}(\bm{k})- i\Delta)^{-1}$ are the retarded and advanced Green's functions, respectively.
Here, $\Delta=\frac{\hbar}{2\tau}$ is the self-energy due to the impurity scattering.

Second, by expanding Taylor series with respect to $\omega$, we obtain
\begin{align}
  K_{(i,\mu)}^\lambda(\omega)= &-\int_{-\infty}^{\infty} \frac{d\xi}{2\pi i} \int\smrm{BZ}\frac{d\bm{k}}{(2\pi)^d}\nonumber \\
  & \ \  \times \big[f_0(\xi) \mathrm{Tr}[j_{\bm{k}}^\lambda G_{\bm{k}}^R(\xi) (\Pi_i\sigma_\mu)G_{\bm{k}}^R(\xi)]\nonumber \\
  &\quad -f_0(\xi)\mathrm{Tr}[j_{\bm{k}}^\lambda G_{\bm{k}}^A(\xi) (\Pi_i\sigma_\mu)G_{\bm{k}}^A(\xi)]\big]\nonumber \\
  &-\hbar\omega \int_{-\infty}^{\infty} \frac{d\xi}{2\pi i} \int\smrm{BZ}\frac{d\bm{k}}{(2\pi)^d}\nonumber \\
  &\ \  \times \big[f_0'(\xi)  \mathrm{Tr}[j_{\bm{k}}^\lambda G_{\bm{k}}^R(\xi) (\Pi_i\sigma_\mu)G_{\bm{k}}^A(\xi)]\nonumber \\
  & \quad  -f_0(\xi) \mathrm{Tr}[j_{\bm{k}}^\lambda (G_{\bm{k}}^R(\xi))^2 (\Pi_i\sigma_\mu)G_{\bm{k}}^R(\xi)]\nonumber \\
  &\quad -f_0(\xi)\mathrm{Tr}[j_{\bm{k}}^\lambda G_{\bm{k}}^A(\xi) (\Pi_i\sigma_\mu)(G_{\bm{k}}^A(\xi))^2]\big]\nonumber \\
  &+O(\omega^2).\label{eqApp:Kexpansion}
\end{align}
As for the first term (i.e. $K_{(i,\mu)}^\lambda(0)$), by performing the analytic continuation $\xi\to \omega_l$, we obtain
\begin{align}
  K_{(i,\mu)}^\lambda(0)=&\frac{1}{\beta}\sum_{\omega_l}\int\smrm{BZ}\frac{d\bm{k}}{(2\pi)^d}\mathrm{Tr}[j_{\bm{k}}^\lambda \mathcal{G}_{\bm{k}}(i\omega_l) (\Pi_i\sigma_\mu)\mathcal{G}_{\bm{k}}(i\omega_l)]\nonumber \\
  =&\frac{1}{\beta}\sum_{\omega_l}\int\smrm{BZ}\frac{d\bm{k}}{(2\pi)^d}\left(-\frac{e}{\hbar}\right)\mathrm{Tr}[\partial_{\bm{k}_\lambda} \mathcal{G}_{\bm{k}}(i\omega_l) (\Pi_i\sigma_\mu)]\nonumber \\
  =&0 \label{eqApp:diamag}.
\end{align}
Here, we use $-\frac{e}{\hbar}\partial_{\bm{k}_\lambda} \mathcal{G}_{\bm{k}}(i\omega_l)=\mathcal{G}_{\bm{k}}(i\omega_l)j_{\bm{k}}^\lambda \mathcal{G}_{\bm{k}}(i\omega_l)$ in the first line and the cancellation of the surface term of $k_\lambda$-integral in the second line.
We note that $K_{(i,\mu)}^\lambda(0)$ also appears as the factor of vertices of the order $A$ i.e. the modulation of the current operator by the Hund coupling and the modulation of the Hund coupling by $A$, which are often referred to as diamagnetic terms.
Since the contributions from vertices of the order $A$ is zero as shown in Eq.~\eqref{eqApp:diamag}, it suffices that we only consider the linear response by the Feynman diagram shown in Fig.~\ref{fig:FeynmanDiagram}.
By diagonaliziing the Green's function and dividing $K_{(i,\mu)}^\lambda(\omega)$ into intra and interband terms, $K_{(i,\mu)}^\lambda(\omega)$ is given as
\begin{widetext}
\begin{align}
  K_{(i,\mu)}^\lambda(\omega)\simeq 
   &-\hbar\omega\int_{-\infty}^{\infty} \frac{d\xi}{2\pi i} \int\smrm{BZ}\frac{d\bm{k}}{(2\pi)^d}\Bigg[\sum_u f_0'(\xi)\mel{u}{j_{\bm{k}}^\lambda}{u}\mel{u}{\Pi_i\sigma_\mu}{u}\frac{1}{\xi-\xi_u+i\Delta}\frac{1}{\xi-\xi_u-i\Delta}\nonumber \\
  &\qquad -\sum_u f_0(\xi) \mel{u}{j_{\bm{k}}^\lambda}{u}\mel{u}{\Pi_i\sigma_\mu}{u}\left(\frac{1}{(\xi-\xi_u+i\Delta)^3}+\frac{1}{(\xi-\xi_u-i\Delta)^3}\right)\nonumber\\
  &\qquad +\sum_{u\neq v} f_0'(\xi)\mel{u}{j_{\bm{k}}^\lambda}{v}\mel{v}{\Pi_i\sigma_\mu}{u}\frac{1}{\xi-\xi_v+i\Delta}\frac{1}{\xi-\xi_u-i\Delta}\nonumber\\
  &\qquad -\sum_{u\neq v} f_0(\xi) \mel{u}{j_{\bm{k}}^\lambda}{v}\mel{v}{\Pi_i\sigma_\mu}{u}\left(\frac{1}{(\xi-\xi_v+i\Delta)^2}\frac{1}{\xi-\xi_u+i\Delta}+\frac{1}{\xi-\xi_v-i\Delta}\frac{1}{(\xi-\xi_u-i\Delta)^2}\right)
  \Bigg]\nonumber \\
  =&-\hbar\omega\int_{-\infty}^{\infty} \frac{d\xi}{2\pi i} \int\smrm{BZ}\frac{d\bm{k}}{(2\pi)^d}\nonumber\\
  &\qquad \times \Bigg[\sum_u f_0'(\xi)\mel{u}{j_{\bm{k}}^\lambda}{u}\mel{u}{\Pi_i\sigma_\mu}{u}\frac{1}{2\Delta}\left(\frac{2\Delta}{(\xi-\xi_u)^2+\Delta^2}+2\frac{2\Delta}{(\xi-\xi_u)^2+\Delta^2}-2\frac{4\Delta^3}{((\xi-\xi_u)^2+\Delta^2)^2}\right)\nonumber\\
  &\qquad +\sum_{u\neq v} f_0(\xi) \mel{u}{j_{\bm{k}}^\lambda}{v}\mel{v}{\Pi_i\sigma_\mu}{u}\left(-\frac{1}{(\xi-\xi_v+i\Delta)^2}\frac{2i\Delta}{(\xi-\xi_u)^2+\Delta^2}+\frac{2i\Delta}{(\xi-\xi_v)^2+\Delta^2}\frac{1}{(\xi-\xi_u-i\Delta)^2}\right)
  \Bigg]
\end{align}

Finally, by using 
\begin{align}
  \frac{2\Delta}{(\xi-\xi_u)^2+\Delta^2}\sim2\pi \delta(\xi-\xi_u),\\
  \frac{4\Delta^3}{((\xi-\xi_u)^2+\Delta^2)^2}\sim2\pi \delta(\xi-\xi_u)
\end{align}
in the weak scattering limit ($\Delta\to 0$), we obtain,
\begin{align}
  K_{(i,\mu)}^\lambda(\omega)= &i\hbar\omega\int\smrm{BZ}\frac{d\bm{k}}{(2\pi)^d}\sum_u f_0'(\xi_u)\mel{u}{j_{\bm{k}}^\lambda}{u}\mel{u}{\Pi_i\sigma_\mu}{u}\frac{1}{2\Delta}\nonumber\\
  &-\hbar\omega\int\smrm{BZ}\frac{d\bm{k}}{(2\pi)^d}\sum_{u\neq v}  \mel{u}{j_{\bm{k}}^\lambda}{v}\mel{v}{\Pi_i\sigma_\mu}{u}\left(\frac{f_0(\xi_v)-f_0(\xi_u)}{(\xi_v-\xi_u)^2}\right).
\end{align}
\end{widetext}
By substituting $\Delta=\frac{\hbar}{2\tau}$ and using the Hermicity of $j_{\bm{k}}^\lambda$ and $\Pi_i\sigma_\mu$, we obtain Eq.~\eqref{eq:sigmaIMuWithC}.
As for the magnon operator formalism, we obtain Eqs.~\eqref{eqApp:CO} and \eqref{eqApp:CI} by the similar calculation.
However, the interband terms of the magnon operator formalism (Eq.~\eqref{eqApp:CI}) cannot be expressed as the imaginary part of matrix elements as in Eq.~\eqref{eq:CI} because $\mathcal{M}^{\mu'}$ is not Hermitian.

\section{Emergent electric field}\label{app:emergentE}
In this section, we derive Eq.~\eqref{eq:emergentEtoContinuum}.
First, we transform $\braket{\bm n_j}{\bm n_{j+1}}$ to the expression in terms of the average and difference of the angles $\theta_j,\phi_j$,
\begin{align}
 & \braket{\bm n_j}{\bm n_{j+1}} \nonumber \\ =&\cos \frac{\theta_j}{2}\cos \frac{\theta_{j+1}}{2}e^{i\frac{\phi_j-\phi_{j+1}}{2}}+\sin \frac{\theta_j}{2}\sin \frac{\theta_{j+1}}{2}e^{-i\frac{\phi_j-\phi_{j+1}}{2}}\nonumber \\
  = & \frac{1}{2}(\cos \overline{\theta}_j+\cos d\theta_j)e^{-id\phi_j}+\frac{1}{2}(-\cos \overline{\theta}_j+\cos d\theta_j)e^{id\phi_j}\nonumber \\
  = & -i\cos \overline{\theta}_j\sin d\phi_j+\cos d\theta_j\cos d\phi_j,
\end{align}
where $\overline{\theta}_j=(\theta_j+\theta_{j+1})/2,d\theta_j=(\theta_{j+1}-\theta_{j})/2,d\phi_j=(\phi_{j+1}-\phi_{j})/2$.
Next, we evaluate the time derivative of the Berry phase in real space using the chain rule,
\begin{align}
  \partial_t\mathrm{Im} \log  \prod_j \braket{\bm n_j}{\bm n_{j+1}}
  = & \sum_j\partial_t\mathrm{Im} \log   \braket{\bm n_j}{\bm n_{j+1}\nonumber } \\
  = & \sum_j[\dot{\theta}_j \mathrm{Im}\partial_{\theta_j}\log\braket{\bm n_j}{\bm n_{j+1}}\nonumber \\
 & \quad +\dot{\theta}_j \mathrm{Im}\partial_{\theta_j}\log\braket{\bm n_{j-1}}{\bm n_{j}}\nonumber \\
 & \quad+\dot{\phi}_j \mathrm{Im}\partial_{\phi_j}\log\braket{\bm n_j}{\bm n_{j+1}}\nonumber \\
 & \quad+\dot{\phi}_j \mathrm{Im}\partial_{\phi_j}\log\braket{\bm n_{j-1}}{\bm n_{j}}]
\end{align}
Using $\partial_{\theta_j}\overline{\theta}_j=\partial_{\theta_j}\overline{\theta}_{j-1}=1/2,\partial_{\theta_j}d\theta_j=-\partial_{\theta_j}d\theta_{j-1}=-1/2,\partial_{\phi_j}d\phi_j=-\partial_{\phi_j}d\phi_{j-1}=-1/2$, we obtain
\begin{widetext}
  \begin{align}
    \mathrm{Im}\partial_{\theta_j}\log\braket{\bm n_j}{\bm n_{j+1}}
    = & \frac{1}{2}\mathrm{Im}\frac{i\sin \overline{\theta}_j\sin d\phi_j+\sin d\theta_j \cos d\phi_j}{-i\cos \overline{\theta}_j\sin d\phi_j+\cos d\theta_j\cos d\phi_j}\nonumber \\
    = & \frac{1}{2}\frac{\sin \overline{\theta}_j\sin d\phi_j\cos d\theta_j\cos d\phi_j+\sin d\theta_j \cos d\phi_j\cos \overline{\theta}_j\sin d\phi_j}{\cos^2 \overline{\theta}_j\sin^2 d\phi_j+\cos^2 d\theta_j\cos^2 d\phi_j} \\
    \mathrm{Im}\partial_{\phi_j}\log\braket{\bm n_j}{\bm n_{j+1}}
    = & \frac{1}{2}\mathrm{Im}\frac{i\cos \overline{\theta}_j\cos d\phi_j+\sin \overline{\theta}_j\sin d\phi_j}{-i\cos \overline{\theta}_j\sin d\phi_j+\cos d\theta_j\cos d\phi_j}\nonumber \\
    = & \frac{1}{2}\frac{\cos \overline{\theta}_j\cos d\phi_j\cos d\theta_j\cos d\phi_j+\sin \overline{\theta}_j\sin d\phi_j\cos \overline{\theta}_j\sin d\phi_j}{\cos^2 \overline{\theta}_j\sin^2 d\phi_j+\cos^2 d\theta_j\cos^2 d\phi_j} \\
    \mathrm{Im}\partial_{\theta_j}\log\braket{\bm n_{j-1}}{\bm n_{j}}
    = & \frac{1}{2}\frac{\sin \overline{\theta}_{j-1}\sin d\phi_{j-1}\cos d\theta_{j-1}\cos d\phi_{j-1}-\sin d\theta_{j-1} \cos d\phi_{j-1}\cos \overline{\theta}_{j-1}\sin d\phi_{j-1}}{\cos^2 \overline{\theta}_{j-1}\sin^2 d\phi_{j-1}+\cos^2 d\theta_{j-1}\cos^2 d\phi_{j-1}} \\
    \mathrm{Im}\partial_{\phi_j}\log\braket{\bm n_{j-1}}{\bm n_{j}}
    = & -\frac{1}{2}\frac{\cos \overline{\theta}_{j-1}\cos d\phi_{j-1}\cos d\theta_{j-1}\cos d\phi_{j-1}+\sin \overline{\theta}_{j-1}\sin d\phi_{j-1}\cos \overline{\theta}_{j-1}\sin d\phi_{j-1}}{\cos^2 \overline{\theta}_{j-1}\sin^2 d\phi_{j-1}+\cos^2 d\theta_{j-1}\cos^2 d\phi_{j-1}}
  \end{align}
\end{widetext}
Considering the continuum limit, we drop $O(d\theta_j^2,d\phi_j^2,d\theta_jd\phi_j)$ terms and obtain
\begin{align}
  \mathrm{Im}\partial_{\theta_j}\log\braket{\bm n_j}{\bm n_{j+1}}
 & \to \frac{1}{2}\sin \overline{\theta}_j d\phi_j \\
  \mathrm{Im}\partial_{\phi_j}\log\braket{\bm n_j}{\bm n_{j+1}}
 & \to \frac{1}{2}\cos \overline{\theta}_j \\
  \mathrm{Im}\partial_{\theta_j}\log\braket{\bm n_{j-1}}{\bm n_{j}}
 & \to \frac{1}{2}\sin \overline{\theta}_{j-1} d\phi_{j-1} \\
  \mathrm{Im}\partial_{\phi_j}\log\braket{\bm n_{j-1}}{\bm n_{j}}
 & \to -\frac{1}{2}\cos \overline{\theta}_{j-1}
\end{align}
Therefore, we obtain the emergent electric field in the continuum limit (Eq.~\eqref{eq:emergentEtoContinuum}) as
\begin{align}
 & \frac{\hbar}{2e|\bm{r}_n-\bm{r}_1|}\sum_j [\dot{\theta}_j (\sin \overline{\theta}_j d\phi_j+\sin \overline{\theta}_{j-1} d\phi_{j-1})\nonumber \\
 & \qquad \qquad \qquad +\dot{\phi}_j (\cos \overline{\theta}_j -\cos \overline{\theta}_{j-1} )]\nonumber \\
  \to & \frac{\hbar}{2e|\bm{r}_n-\bm{r}_1|}\int_{r_1}^{r_n}\sin \theta (\dot{\theta}\partial_x \phi -\dot{\phi}\partial_x \theta) dx.
\end{align}
Here, we use $\cos \theta_j-\cos \theta_{j-1}\to -\sin \theta_j d\theta_{j-1},\overline{\theta}_j\to \theta$, and  $d\phi_j\to (\phi /2) dx$.

\section{numerical calculation with spin-orbit coupling}\label{app:spinorbit}

\begin{figure}
  \includegraphics[width=\linewidth]{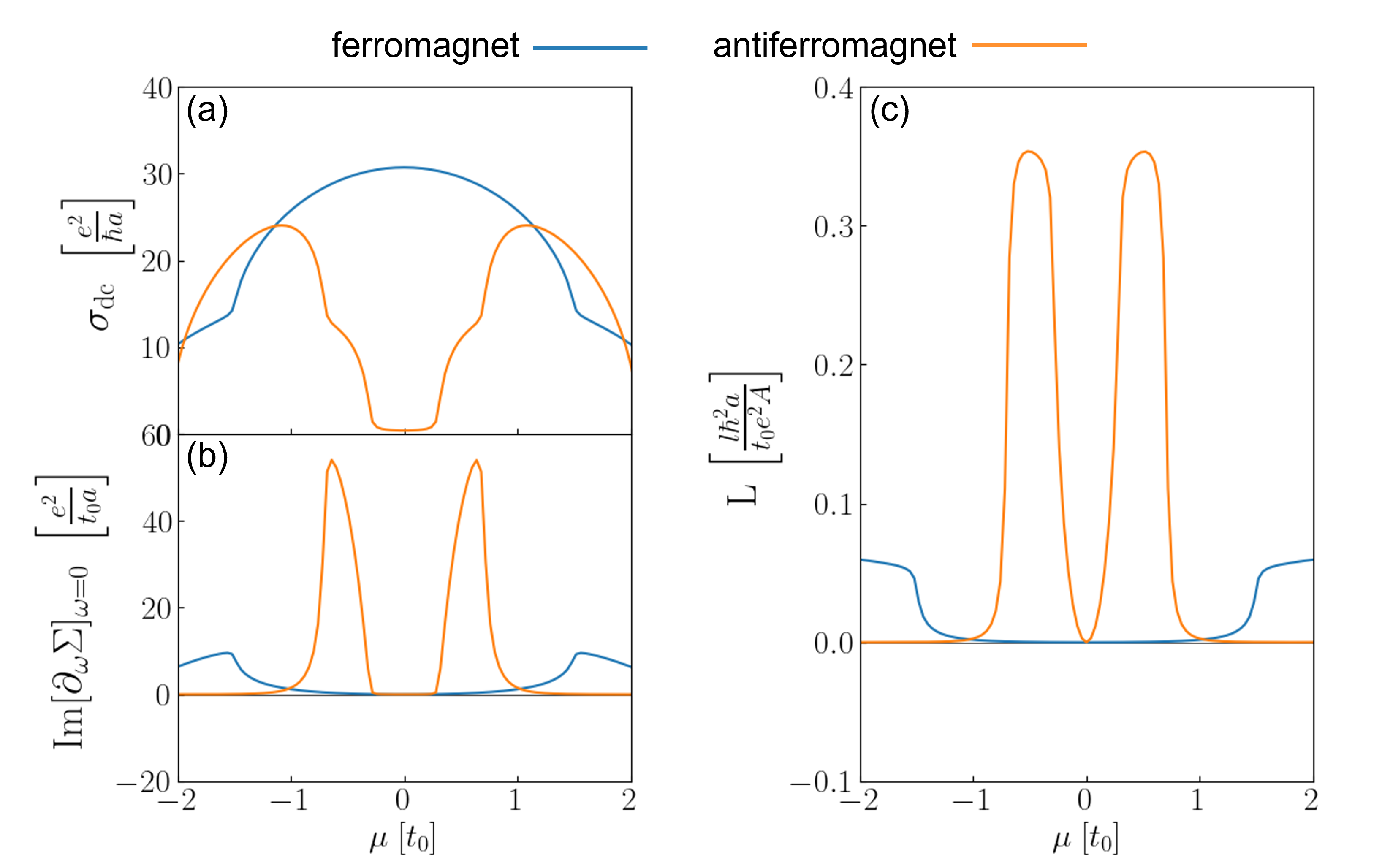}
  \caption{(a) DC conductivity $\sigma\smrm{dc}$, (b) the imaginary part of the frequency derivative of the conductivity mediated by magnon $\mathrm{Im}[\partial_\omega \Sigma] _{\omega=0}$ and (c) the inductance $L$, as a function of the chemical potential $\mu$ for the ferromagnetic state with SOC (blue) and the antiferromagnetic state with SOC (orange). We adopt the same parameter set as Fig.~\ref{fig:Inductance} and the SOC strength $\bm{\lambda}=(0.1t_0,0,0)$.}
  \label{fig:InductanceSOC}
\end{figure}

In this section, we present the numerical results for the inductance of a ferromagnet and an antiferromagnet in the presence of spin-orbit coupling (SOC).
While the SOC induced emergent inductance is shown in previous theoretical works~\cite{Yamane2022,Araki2023}, here we demonstrate that the present method can also be applied to the model with SOC.
We apply the present method to the following 1D models with SOC: ferromagnetic state and antiferromagnetic state.
The Hamiltonians of the itinerant electron and Hund coupling of both models are given as
\begin{align}
  H\smrm{itinerant}\uprm{ferro(antiferro)}=&\sum_{i}[-t_0c_{i}^\dagger c_{i+1}+h.c.]\nonumber \\
  &\qquad+\sum_i[-c_{i}^\dagger i\bm{\lambda}\cdot \bm{\sigma} c_
  {i+1}+h.c.], \\
  H\smrm{Hund}=&-\sum_{i}\frac{J\uprm{H}}{2}\bm{S}_i \cdot c_{i}^\dagger\bm{\sigma}c_{i},
\end{align} 
where the SOC is introduced with $\bm{\lambda}$.
The spin Hamiltonian of each model is given as
\begin{align}
  H\smrm{m}\uprm{ferro}=&\sum_{i}-J\smrm{ex}\bm{S}_i\cdot \bm{S}_{i+1}\\
  H\smrm{m}\uprm{antiferro}=&\sum_{i}J\smrm{ex}\bm{S}_i\cdot \bm{S}_{i+1}
\end{align}
By performing the orthogonal transformation Eq.~\eqref{eq:T}, we obtain $\tilde{J}_{j,\tau}$ as,
\begin{align}
  \begin{array}{c}
    \tilde{J}_{j,1}^{xx}=\pm J\smrm{ex}, \\ \tilde{J}_{j,1}^{yy}=\tilde{J}_{j,1}^{zz}=J\smrm{ex},\
    \tilde{J}_{j,0}^{zz}=p.
  \end{array} \label{eq:tildeJferro}
\end{align}
Here, the sign of $\tilde{J}_{j,1}^{xx}$ corresponds to the ferromagnetic state ($+$) and the antiferromagnetic state ($-$).

Figure~\ref{fig:InductanceSOC} shows the DC conductivity $\sigma\smrm{dc}$, the imaginary part of the frequency derivative of the conductivity mediated by magnon $\mathrm{Im}[\partial_\omega \Sigma] _{\omega=0}$ and the inductance $L$, as a function of the chemical potential $\mu$ for the ferromagnetic state with SOC and the antiferromagnetic state with SOC.
As for the ferromagnetic state with SOC, $\sigma\smrm{dc}$, $\mathrm{Im}[\partial_\omega \Sigma] _{\omega=0}$ and $L$ show qualitatively the same behavior as the cases of helical state without SOC.
This shows that the SOC and the helical spin structure play essentially the same role in generating emergent inductance as shown in Eq.~\eqref{eq:SOCchange}.
Note that quantitative difference arises because of the modulation of effective hopping $t_{j,\tau}\to \sqrt{t_{j,\tau}^2+\lambda^2}$ (Eq.~\eqref{eq:SOCchange}) and the absence of DM interaction in the spin Hamiltonian (compare Eq.~\eqref{eq:tildeJhelical} and Eq.~\eqref{eq:tildeJferro}).
In the antiferromagnetic case, the spin splitting of energy bands appears in the vicinity of the band gap ($\mu\simeq 0$). Thus the emergent inductance also becomes large near the band gap.

\bibliographystyle{apsrev4-1}
\bibliography{references}

\end{document}